\documentclass[12pt]{article}

\usepackage{mathrsfs}
\usepackage[T1]{fontenc}
\usepackage{mathpazo}
\usepackage{setspace}
\usepackage{amsfonts}
\usepackage{amssymb}
\usepackage{amsmath}
\usepackage{esint}                   
\usepackage{epsfig}
\usepackage{latexsym}
\usepackage{color}
\usepackage{graphicx}
\usepackage{hyperref}
\usepackage{nicefrac}
\usepackage[latin1]{inputenc}
\usepackage{pstricks}
\usepackage{slashed}
\usepackage{multirow}
\usepackage{ulem}
\usepackage{comment}
\usepackage[nosort]{cite}
\usepackage{datetime}
\usepackage{tikz-cd}
\usepackage{float}

\usepackage{comment}
\usepackage{ytableau}

\def\hybrid{\topmargin -30pt    \oddsidemargin 0pt 
        \headheight 0pt \headsep 0pt
        \textwidth 6.25in       
        \textheight 9.5in       
        \marginparwidth .875in
        \parskip 5pt plus 1pt   \jot = 1.5ex}

\hybrid

\def\baselinestretch{1.2}

\catcode`\@=11

\def\marginnote#1{}
\newcount\hour
\newcount\minute
\newtoks\amorpm
\hour=\time\divide\hour by60
\minute=\time{\multiply\hour by60 \global\advance\minute by-\hour}
\edef\standardtime{{\ifnum\hour<12 \global\amorpm={am}%
        \else\global\amorpm={pm}\advance\hour by-12 \fi
        \ifnum\hour=0 \hour=12 \fi
        \number\hour:\ifnum\minute<10 0\fi\number\minute\the\amorpm}}
\edef\militarytime{\number\hour:\ifnum\minute<10 0\fi\number\minute}

\def\draftlabel#1{{\@bsphack\if@filesw {\let\thepage\relax
   \xdef\@gtempa{\write\@auxout{\string
      \newlabel{#1}{{\@currentlabel}{\thepage}}}}}\@gtempa
   \if@nobreak \ifvmode\nobreak\fi\fi\fi\@esphack}
        \gdef\@eqnlabel{#1}}
\def\@eqnlabel{}
\def\@vacuum{}
\def\draftmarginnote#1{\marginpar{\raggedright\scriptsize\tt#1}}

\def\draft{\oddsidemargin -.5truein
        \def\@oddfoot{\sl preliminary draft \hfil
        \rm\thepage\hfil\sl\today\quad\militarytime}
        \let\@evenfoot\@oddfoot \overfullrule 3pt
        \let\label=\draftlabel
        \let\marginnote=\draftmarginnote
   \def\@eqnnum{(\theequation)\rlap{\kern\marginparsep\tt\@eqnlabel}%
\global\let\@eqnlabel\@vacuum}  }

\def\draft2{
        \def\@oddfoot{\sl preliminary draft \hfil
        \rm\thepage\hfil\sl\today\quad\militarytime}
        \let\@evenfoot\@oddfoot \overfullrule 3pt
        \let\marginnote=\draftmarginnote
   \def\@eqnnum{(\theequation)\rlap{\kern\marginparsep\tt\@eqnlabel}%
\global\let\@eqnlabel\@vacuum}  }

\def\preprint{\twocolumn\sloppy\flushbottom\parindent 2em
        \leftmargini 2em\leftmarginv .5em\leftmarginvi .5em
        \oddsidemargin -.5in    \evensidemargin -.5in
        \columnsep .4in \footheight 0pt
        \textwidth 10.in        \topmargin  -.4in
        \headheight 12pt \topskip .4in
        \textheight 6.9in \footskip 0pt
        \def\@oddhead{\thepage\hfil\addtocounter{page}{1}\thepage}
        \let\@evenhead\@oddhead \def\@oddfoot{} \def\@evenfoot{} }

\def\numberbysection{\@addtoreset{equation}{section}
        \def\theequation{\thesection.\arabic{equation}}}

\def\underline#1{\relax\ifmmode\@@underline#1\else
        $\@@underline{\hbox{#1}}$\relax\fi}

\def\titlepage{\@restonecolfalse\if@twocolumn\@restonecoltrue\onecolumn
     \else \newpage \fi \thispagestyle{empty}\c@page\z@
        \def\thefootnote{\fnsymbol{footnote}} }

\def\endtitlepage{\if@restonecol\twocolumn \else \newpage \fi
        \def\thefootnote{\arabic{footnote}}
        \setcounter{footnote}{0}}  

\catcode`@=12
\relax

\def\figcap{\section*{Figure Captions\markboth
        {FIGURECAPTIONS}{FIGURECAPTIONS}}\list
        {Figure \arabic{enumi}:\hfill}{\settowidth\labelwidth{Figure
999:}
        \leftmargin\labelwidth
        \advance\leftmargin\labelsep\usecounter{enumi}}}
 \relax
\def\tablecap{\section*{Table Captions\markboth
        {TABLECAPTIONS}{TABLECAPTIONS}}\list
        {Table \arabic{enumi}:\hfill}{\settowidth\labelwidth{Table
999:}
        \leftmargin\labelwidth
        \advance\leftmargin\labelsep\usecounter{enumi}}}
 \relax
\def\reflist{\section*{References\markboth
        {REFLIST}{REFLIST}}\list
        {[\arabic{enumi}]\hfill}{\settowidth\labelwidth{[999]}
        \leftmargin\labelwidth
        \advance\leftmargin\labelsep\usecounter{enumi}}}
 \relax
\makeatletter
\newcounter{pubctr}
\def\publist{\@ifnextchar[{\@publist}{\@@publist}}
\def\@publist[#1]{\list
        {[\arabic{pubctr}]\hfill}{\settowidth\labelwidth{[999]}
        \leftmargin\labelwidth
        \advance\leftmargin\labelsep
        \@nmbrlisttrue\def\@listctr{pubctr}
        \setcounter{pubctr}{#1}\addtocounter{pubctr}{-1}}}
\def\@@publist{\list
        {[\arabic{pubctr}]\hfill}{\settowidth\labelwidth{[999]}
        \leftmargin\labelwidth
        \advance\leftmargin\labelsep
        \@nmbrlisttrue\def\@listctr{pubctr}}}
 \relax
\makeatother

\def\be{\begin{equation}}
\def\ee{\end{equation}}
\def\ba{\begin{eqnarray}}
\def\ea{\end{eqnarray}}

\def\del{\partial}

\def\bx {{\bf x}}
\def\bk {{\bf k}}
\def\bw {{\bf w}}

\def\d{\delta}
\def\D{\Delta}
\def\e{\epsilon}

\def\om{\omega}

\def\l{\lambda}
\def\L{\Lambda}

\def\uu{\vskip -0.2cm}

\def\cM{{\cal M}}

\def\no{\noindent}

\def\qq{\qquad}

\def\IR{\relax{\rm I\kern-.18em R}}

\def\bx {{\bf x}}
\def\bw {{\bf w}}
\def\bu {{\bf u}}
\def\bv {{\bf v}}

\def\bA {{\bf a}}
\def\bB {{\bf b}}
\def\ab {{\bf a} \cdot {\bf b}}
\def\eo {\varepsilon_1}
\def\et {\varepsilon_2}

\def\inv{^{\raise.0ex\hbox{${\scriptscriptstyle -}$}\kern-.05em 1}}

\def \ha {{\frac{1}{2}}}

\def \ov {\over}

\def\diag{{\rm diag}}

\def\half{{\textstyle {1 \over 2}}}

\newcommand{\bb}{\hskip -0.1cm}
\newcommand{\hb}{\hskip -0.05cm}
\newcommand{\ssb}{\hskip -0.03cm}
\newcommand{\hp}{\hskip -0.05cm + \hskip -0.05cm}
\newcommand{\hm}{\hskip -0.05cm - \hskip -0.05cm}

\def\tr{\textrm{Tr}}

\begin{document}


\renewcommand{\theequation}{\thesection.\arabic{equation}}
\csname @addtoreset\endcsname{equation}{section}

\begin{titlepage}
\begin{center}

\renewcommand*{\thefootnote}{\arabic{footnote}}

\phantom{xx}
\vskip 0.5in

{\large {\bf Ferromagnets from higher $SU(N)$ representations}}

\vskip 0.5in

{\bf Alexios P. Polychronakos$^{1,2}$}\hskip .15cm and \hskip .15cm
{\bf Konstantinos Sfetsos}$^{3}$

\vskip 0.14in

${}^1\!$ Physics Department, the City College of New York\\
160 Convent Avenue, New York, NY 10031, USA\\
{\footnotesize{\tt apolychronakos@ccny.cuny.edu}}\\
\vskip 0.3cm
${}^2\!$ The Graduate School and University Center, City University of New York\\
365 Fifth Avenue, New York, NY 10016, USA\\
{\footnotesize{\tt apolychronakos@gc.cuny.edu}}

\vskip .15in

${}^3\!$
Department of Nuclear and Particle Physics, \\
Faculty of Physics, National and Kapodistrian University of Athens, \\
Athens 15784, Greece\\
{\footnotesize{ ksfetsos@phys.uoa.gr}}\\

\vskip .3in
\today

\vskip .2in

\end{center}

\vskip .2in

\centerline{\bf Abstract}

\no
We present a general formalism for deriving the thermodynamics of ferromagnets consisting of
"atoms" carrying an arbitrary irreducible representation of $SU(N)$ and coupled through long-range
two-body quadratic interactions. Using this formalism, we derive the thermodynamics and phase
structure of ferromagnets
with atoms in the doubly symmetric or doubly antisymmetric irreducible representations.
The symmetric representation leads to a paramagnetic and a ferromagnetic phase with transitions
similar to the ones for the fundamental representation studied before. The antisymmetric representation
presents qualitatively new features, leading to a paramagnetic and two distinct ferromagnetic phases
that can coexist over a range of temperatures, two of them becoming metastable. Our results are relevant
to magnetic systems of atoms with reduced symmetry in their interactions compared to the fundamental case.

\vskip .4in

\vfill

\end{titlepage}
\vfill
\eject



\def\baselinestretch{1.2}
\baselineskip 20 pt

\newcommand{\eqn}[1]{(\ref{#1})}

\tableofcontents


\section{Introduction}
\label{intro}
Magnetic systems with higher internal $SU(N)$ symmetry are enjoying a revival of interest in physics,
both experimental and theoretical.
Such systems have been considered in the context of ultracold atoms \cite{Dud,Ghu,Gor,Zha,Mag,Cap,Mukherjee:2024ffz}, spin chains \cite{Aff,Pola},
and interacting atoms on lattice cites \cite{KT,BSL,RoLa,YSMOF,TK,Totsuka,TK2}.
They were also studied in the presence of external $SU(N)$ magnetic fields \cite{DY,YM,HM}.
The $SU(N)$ symmetry in cold atoms such as ${}^{137}$Yb or
${}^{87}$Sr emerges through the virtual independence of atom interactions on the nuclear spin
(hyperfine structure) of the atoms $s$, the $N = 2s+1$ nuclear spin states behaving as states in the
fundamental representation of $SU(N)$, and can be further enhanced through the
near-degeneracy of low-lying thermally occupied states. Quasiparticle excitations in many-body
fermionic atom systems carry $SU(N)$ degrees of freedom and their effective interaction leads to
nonabelian magnetism and novel collective effects.

In recent work \cite{PSferro,PSlargeN,PSferroN} we considered ferromagnets consisting of atoms on fixed positions
with $SU(N)$ degrees of freedom and pairwise interactions, a setting differing from that in cold atoms, where
the dynamical effect of $SU(N)$ degrees of freedom emerges through quasiparticle interactions.
We derived the thermodynamics of such systems \cite{PSferro} and uncovered an intricate and
nontrivial phase
structure with qualitatively new features compared to standard $SU(2)$ ferromagnets. In particular,
at zero magnetic field the system has three critical temperatures (vs.\ only one Curie temperature
for $SU(2)$) and a regime where the paramagnetic and ferromagnetic phases become metastable.
Spontaneous breaking of the global $SU(N)$ symmetry arises in the
$SU(N) \to SU(N-1) \times U(1)$ channel at zero external magnetic field, and generalizes to other
channels in the presence of non-Abelian magnetic fields. Due to the presence of metastable states,
the $SU(N)$ system exhibits hysteresis, both in the magnetic field and in the temperature.
We also examined the ferromagnet in the limit where the rank of the group $N$ scales as the square root
of the number of atoms \cite{PSlargeN,PSferroN}, and uncovered an even more intricate phase structure,
with the paramagnetic phase splitting into two distinct phases, a triple critical point, and two different
temperature scales.

All previous work, including \cite{PSferro,PSlargeN,PSferroN}, considered atoms carrying the fundamental representation of
$SU(N)$; that is, $N$ states fully mixing under the action of the symmetry group. It is of interest to also
consider situations in which the states of each atom transform in an {\it arbitrary} irreducible representation
(irrep) $\chi$ of $SU(N)$, of dimension $d_\chi$ higher than that of the fundamental. This, apart from its
theoretical interest, is also phenomenologically relevant. The case of atoms in the fundamental irrep
corresponds to maximal symmetry, and the two-atom interactions are essentially of the exchange type.
Higher irreps $\chi$ correspond to situations with a {\it reduced} symmetry: two-atom interactions are still
$SU(N)$-invariant, but the $d_\chi$ states per atom are not arbitrarily mixed under the action of the
symmetry group $SU(N)$. Higher irreps can also probe situations with smaller, or vanishing, polarization
for the atoms. For example, the adjoint irrep of $SU(N)$ is self-conjugate, and thus the atoms have no
polarization. It is on interest to uncover the phase structure and symmetry breaking patterns of such
ferromagnets.

In this paper we study $SU(N)$ ferromagnetism for atoms carrying an arbitrary irrep $\chi$ per atom.
Although the mathematical complexity of the situation increases, we were able to explicitly perform the
analysis in the thermodynamic limit and derive equilibrium equations for the states of the system.
Examination of a couple of higher irreps reveals the existence of additional phases and corresponding
symmetry breaking patterns as a function of the temperature, with several of them coexisting as
metastable states. Transitions are typically discontinuous (in energy) or first order (in free energy)
for generic $N$, with special situations arising for small $N$ for each irrep.

\no
The organization of the paper is as follows:
In section \ref{model} we introduce the model of atoms with arbitrary $SU(N)$ irrep $\chi$ and review
the essential group theory facts required for its analysis. We then present a general method for deriving its equilibrium equations in the thermodynamic limit, and derive the conditions for stability of its configurations.
In section \ref{fundamental} we apply the general method of this paper to analyze the case of the
fundamental irrep, review its properties, and make contact with our previous results  \cite{PSferro}.
In section \ref{symmetric} we analyze the doubly symmetric irrep and derive its thermodynamics, showing
that it has properties qualitatively similar to those of the fundamental. We also examine its large-temperature
and large-$N$ limits.
In section \ref{antisymmetric} we study the doubly antisymmetric irrep and derive its
thermodynamics. We uncover an additional phase and a more intricate phase transition structure, with
qualitatively new features compared to the fundamental and symmetric irreps. We also examine the
cases $N=2,3,4$, in which the model reduces or acquires special properties, and
derive its large-temperature and large-$N$ limits.
In section \ref{conclusions} we present our conclusions, as well as some speculations and
directions for possible future work.
Finally, in the Appendix we derive the general stability conditions for systems of
the type appearing in this paper, which is by itself an interesting mathematical topic.

\section{The model and the thermodynamic limit}
\label{model}
  
In this section we first review the essential features of the model introduced in \cite{PSferro}.
We consider a set of $n$ atoms at fixed positions, each carrying an irreducible
representation (irrep) $\chi$ of $SU(N)$ and interacting with
ferromagnetic-type interactions. Denoting by $j_{r,a}$ the $d_\chi\times d_\chi$-dimensional generators of
$SU(N)$ in the irrep $\chi$ acting on atom $r$ at position $\vec r$,
the interaction Hamiltonian of the full system is
\be
H_0 = \sum_{r,s =1}^n c_{{\vec r},{\vec s}} \sum_{a= 1}^{N^2 -1} j_{r,a} \, j_{s,a}\ ,
\ee
where $c_{{\vec r},{\vec s}} = c_{{\vec s},{\vec r}}$ is the strength of the interaction between atoms $r$ and $s$.
This Hamiltonian involves an isotropic quadratic coupling between the generators in irrep $\chi$ of
the $n$ commuting $SU(N)$ groups of the atoms. 
Assuming translation invariance, $c_{{\vec r},{\vec s}} = c_{{\vec r}-{\vec s}}$, and also that the mean-field
approximation is valid,\footnote{The validity of the mean field approximation is
strongest in three dimensions, since every atom has a higher number of near neighbors and the statistical
fluctuations of their averaged coupling are weaker, but is expected to also hold in lower dimensions.}
each atom will interact with the average of the $SU(N)$ generators of the remaining atoms; that is,
\be
\sum_{{\vec r},{\vec s}} c_{{\vec r}-{\vec s}} \, j_{{\vec r},a} \, j_{{\vec s},a} = \sum_{\vec r} j_{{\vec r},a}
\sum_{\vec s} c_{\vec s}\, j_{{\vec r}+{\vec s},a}  \simeq \sum_r j_{r,a}
\Bigl(\sum_{{\vec s}} c_{\vec s} \Bigr)\,
{1\over n} \sum_{s'=1}^n j_{s',a} = -{c \over n} J_a \, J_a\ ,
\label{meanfield}\ee
where we defined the total $SU(N)$ generator\uu\uu
\be
J_a = \sum_{s=1}^n j_{s,a}
\ee
and the effective mean coupling\uu\uu\uu
\be
c = - \sum_{{\vec s}} c_{\vec s}\ .
\ee
The minus sign is introduced such that ferromagnetic interactions, driving atom states to align,
correspond to positive $c$. Altogether, the effective interaction is proportional to the quadratic Casimir
of the total $SU(N)$ generators
\be
H_0 = -{c \over n} \sum_{a=1}^{N^2 -1} J_a^2\ .
\label{H0}\ee
In the presence of external magnetic fields $B_i$ coupling to the Cartan generators of each atom $h_{s,i}$,
the Hamiltonian acquires the extra term
\be
H_B = \sum_{s=1}^n \sum_{i=1}^N B_i h_{s,i} = \sum_{i=1}^N B_i H_i\ ,
\ee
with $H_i$ the total $SU(N)$ Cartan generators. Altogether, the full Hamiltonian of the model is
\be
H = H_0 + H_B = -{c \over n} \sum_{a=1}^{N^2 -1} J_a^2\  +\sum_{i=1}^N B_i H_i \ .
\label{H}\ee
We can assume that $\displaystyle \sum_{i=1}^N B_i =0$ since $\displaystyle \sum_{j=i}^N H_i =0$
and the $U(1)$ part decouples.

\no
Calculating the partition function involves decomposing the full Hilbert space of the tensor product of $n$
irreps $\chi$ of $SU(N)$ into irreducible representations, each having a fixed quadratic Casimir. This is
most conveniently done in the fermion momentum representation: to each irrep with row lengths
$\ell_1 \geqslant \ell_2 \geqslant \cdots \geqslant \ell_{N-1}$ in its Young tableau (YT)
we map a set of $N$ distinct non-negative integers $k_1 > k_2 > \cdots > k_N$ such that
\be
\label{lengy}
\ell_i = k_i - k_N +i -N \ 
\ee
and we label the irrep with its corresponding vector $\bk = \{k_1,\dots,k_N\}$.
The $k_i$ are, in principle, defined up to an overall common shift $k_i \to k_i + C$ which leaves $\ell_i$ unaffected,
the sum $\sum_i k_i$ representing the $U(1)$ charge of the irrep. However, for our purposes, in which irreps
$\bk$ arise from the decomposition of the tensor product of $n$ irreps $\chi$, it will be convenient
to fix the sum of $k_i$ to the specific value
\be
{\sum_{i=1}^N k_i }= n\;\bb n_{\hskip -0.03cm \chi} +{N(N-1)\over 2}\ ,
\label{lengyt}
\ee
where $n_\chi$ is the number of boxes in the irrep $\chi$ of each of the $n$ atoms.

\noindent
The relevant quantities for our ferromagnetic model are:\\
$\bullet$
The character (trace) in the irrep $\bk$ of an $SU(N)$ element $U = \diag\{z_1,\dots,z_N\}$,
with $z_1 \cdots z_N = 1$, given by
\be
{\bf Tr}_\bk U = {\psi_\bk ({\bf z}) \over \Delta ({\bf z} )}\ ,
\label{tra}\ee
where ${\bf z} = \{z_1,\dots,z_N\}$, $\psi_\bk ({\bf z})$ is the Slater determinant
\be
\label{slat}
\psi_{\bf k}({\bf z} ) = \left|
\begin{array}{ccccccccc}
z_1^{k_1} & z_1^{k_2} & \cdots & z_1^{k_{N-1}}  & z_1^{k_N} \\
z_2^{k_1} & z_2^{k_2} & \cdots & z_2^{k_{N-1}}  & z_2^{k_N} \\
\vdots & \vdots & \ddots & \vdots & \vdots  \\
z_N^{k_1} & z_N^{k_2} & \cdots & z_N^{k_{N-1}} & z_N^{k_N}
\end{array}
\right |\ ,
\ee
which is antisymmetric under the interchange of any two $z_i$ and of any two $k_i$, and $\D({\bf z})$ is
the Vandermonde determinant
\be
\label{vanderm}
\Delta ({\bf  z}) =  \left|
\begin{array}{ccccccccc}
z_1^{N-1} & z_1^{N-2} & \cdots & z_1  & 1 \\
z_2^{N-1} & z_2^{N-2} & \cdots & z_2  & 1 \\
\vdots & \vdots & \ddots & \vdots & \vdots  \\
z_N^{N-1} & z_N^{N-2} & \cdots & z_N  & 1 \\
\end{array}
\right |\ = \prod_{j>i=1}^N (z_i - z_j )\ ,
\ee
which is the Slater determinant (\ref{slat}) for the singlet irrep with $k_i = N-i$.
As a check of (\ref{tra}), the limit $z_i \to 1$ yields
\be
\tr_{\bf k} {\bf 1} = \dim ({\bf k}) = { \D({\bf k})\ov  \displaystyle \prod_{s=1}^{N-1} s!} = \prod_{j>i=1}^N {k_i - k_j  \over j-i}
= \prod_{j>i=1}^N {\ell_i - \ell_j +j-i \over j-i}\ ,
\label{dimk}\ee
which is the standard expression for the dimension of the irrep.

\noindent
$\bullet$ 
The quadratic Casimir $C_2 (\bk)$ of irrep $\bk$, given by
\be
\begin{split}
C_2 (\bk) &  = {1\over 2}\sum_{i=1}^N k_i^2 -{1\over 2N}\left(\sum_{i=1}^N k_i\right)^2 -{N(N^2 -1)\over 24}\cr
&={1\over 2}\sum_{i=1}^N k_i^2 -{1\over 2N}\left({n n_\chi } +{N(N-1)\over 2}\right)^2
-{N(N^2 -1)\over 24}\ .
\label{caso}
\end{split}
\ee
$\bullet$
The multiplicity $d_{n;{\bf k}}$ of the irrep $\bk$ in the decomposition of the product of $n$ irreps $\chi$.
Calling
$\chi({\bf z}) = \tr_\chi (U)$ the character of $\chi$, with $U=\diag\{z_1,\dots,z_N\}$,
$d_{n;{\bf k}}$ satisfies \cite{PScompo}
 \be
 \sum_{\bf k} d_{n;{\bf k}} z_1^{k_1}\dots z_N^{k_N} = \D({\bf z}) \chi^n({\bf z})\  ,
 \ee
 provided that the sum of $k_i$ satisfies the constraint \eqn{lengyt}.
 Thus, $d_{n;{\bf k}}$ obtains from a multiple integration over $z_i$ on the complex plane around the origin as 
 \be
 d_{n;{\bf k}} = {1\ov (2 \pi i)^N} \prod_{i=1}^N   \oint_C  {d z_i\ov z_i^{k_i+1}} \D({\bf z}) \chi^n({\bf z})\ .
 \label{dnz}\ee
 We note that $\chi$ is a homogeneous polynomial in the $z_i$ with degree of homogeneity $n_\chi$, the number
of boxes in the irrep $\chi$. 

\noindent
The partition function of the model in temperature $T = \beta^{-1}$ in the presence of magnetic
 fields $B_i$ can be written, following  \cite{PSferro}, as the explicit sum over $k_i$
 \be
Z= \sum_{\text{states}} e^{-\beta H} =
\sum_{\langle{\bf k}\rangle} d_{n;{\bf k}}\, e^{{\beta c\over n} C^{(2)} ({\bf k})}\,
\tr_{\bf k} \exp\Bigl( \beta \sum_{j=1}^N B_j H_j  \Bigr)\ ,
\ee
where $\langle \bk \rangle$ denotes the set of distinct ordered nonnegative integers $k_i$ satisfying the constraint \eqn{lengyt}.
Since the Cartan generators $H_i$ are diagonal, the magnetic trace can be rewritten as \vskip -0.8cm
\be
\tr_{\bf k} \exp\Bigl( \beta \sum_{j=1}^N B_j H_j  \Bigr) = \tr_{\bf k}\, 
\diag\{ e^{\beta B_1},\dots,e^{\beta B_N} \}\ ,
\ee
for a new set of $B_i$, still satisfying $\sum_{i=1}^N B_i =0$, which are  linear combinations of the previous ones. This trace can be evaluated using \eqn{tra} as
\be
\tr_{\bf k}\, \diag\{ e^{\beta B_i} \} =  {\psi_\bk (e^{\beta{\bf B}}) \over \Delta (e^{\beta{\bf B}})}~,~~~
e^{\beta{\bf B}} = \{ e^{\beta B_1},\dots,e^{\beta B_N} \} \ .
\ee
Using also the results (\ref{caso}-\ref{dnz}) above, and ignoring trivial constants in $C^{(2)}(\bk)$, we obtain
\ba
Z &=& {1\ov (2 \pi i)^N}  \sum_{\langle\bk\rangle} \prod_{i=1}^N \oint {dz_i\ov z_i^{k_i +1} }\D({\bf z}) \chi^n ({\bf z})\,
e^{{\beta c\over 2n}\sum_i k_i^2}\, {\psi_\bk (e^{\beta{\bf B}}) \over \Delta (e^{\beta{\bf B}})} \cr
&=& {1\ov (2 \pi i)^N \D(e^{\beta{\bf B}})}  \sum_{\bk} \prod_{i=1}^N \oint {dz_i\ov z_i^{k_i +1} }\D({\bf z}) \chi^n ({\bf z})\,
e^{{\beta c\over 2n}\sum_i k_i^2 +\beta \sum_i B_i k_i}\ .
\ea
In the second line we made the sum over $k_i$ unrestricted, since the $z_i$-integral is antisymmetric in the $k_i$ due to
the presence of  $\D({\bf z})$, and thus summing over all $k_i$ picks the antisymmetric part of $e^{\beta \sum_i B_i k_i}$,
which reproduces $\psi_\bk (e^{\beta{\bf B}})$. The constraint \eqn{lengyt} on the sum of $k_i$ is also reproduced by the
$z_i$-integral: under the change of variables $z_i \to \alpha z_i$, the character and the Vandermonde factor transform as
\be
\label{chidel}
\chi (\alpha {\bf z}) = \alpha^{n_\chi} \chi({\bf z}) \ ,\qq  \D (\alpha {\bf z}) = \alpha^{N(N-1)/2}\, \D ({\bf z}) 
\ee
and the integral picks up a factor
$\alpha^{n n_\chi +N(N-1)/2 -\sum_i k_i}$ and therefore vanishes whenever \eqn{lengyt} does not hold.
We also note that for $B_i = 0$, both the $z_i$-integral and $\D(e^{\beta{\bf B}})$ vanish. Upon taking the limit
$B_i \to 0$ we recover the formula \eqn{dimk} for the dimension of irrep $\bk$, obtaining
\be
Z = {1\ov (2 \pi i)^N \prod_{s=1}^{N-1} s!}  \sum_{\bk} \prod_{i=1}^N \oint {dz_i\ov z_i^{k_i +1} }\D({\bf z}) \chi^n ({\bf z})\,
\D ({\bk}) \, e^{{\beta c\over 2n}\sum_i k_i^2}\ , ~~~~\text{for}~~ B_i =0\ .
\ee

\noindent
An explicit evaluation of the integrals above, or equivalently of  $d_{n;{\bf k}}$   can, in principle, be done on a case-by-case basis, as demonstrated for the fundamental representation in \cite{PScompo}. However, this procedure yields results that are not particularly illuminating, even for low-dimensional irreps. 
Instead, we are interested in the thermodynamic limit  $n\gg 1$, which can be obtained directly for general irreps using the procedure outlined below.


\no
From \ref{lengyt} we see that in the limit $n\gg 1$ the
$k_i$ scale as $n$. Setting 
\be
k_i = n x_i
\label{kx}
\ee
in the partition function, turning the sum over $k_i$ into an integral over $x_i$, and ignoring subleading in $n$
terms in the exponent, we finally obtain
\be
Z= {n^N\ov (2 \pi i)^N} \prod_{i=1}^N \int dx_i  \oint {dz_i\ov z_i} e^{-n \beta F({\bf x},{\bf z})}\ ,
\ee 
where
\be
\label{ufhju1}
F({\bf x},{\bf z})=  \sum_{i=1}^N \Big( T  x_i \ln z_i- {c x_i^2\ov 2} - B_i x_i\Big) -T \ln \chi({\bf z})\ ,
\ee
represents the free energy per atom in terms of the intensive variables $x_i$.

\subsection{Saddle point equations}

For $n\gg 1$ we can use the saddle point approximation, in which $F({\bf x},{\bf z})$ is minimized both in $z_i$ and $x_i$. The
saddle point equations are
\be
\label{eqsfxz}
\begin{split}
& {\del F\ov \del z_i}= T\bigg( {x_i\ov z_i}  -  {1\ov \chi} {\del \chi \ov \del z_i}\bigg) =0\ ,
\\
& {\del F\ov \del x_i}= T \ln z_i -c x_i - B_i  =0
\end{split}
\ee
and represent the conditions for thermodynamic equilibrium.
Note that even thought the $z_i$ could in principle be complex, reality of the $x_i$ implies that the $z_i$ are real and positive. 
Hence, we define a new set of thermodynamically intensive variables $w_i$ as 
\be
 z_i = e^{w_i}\ .
 \ee 
 Calling 
 \be
 \label{lax}
 \lambda =  \ln \chi\ , 
\ee
the free energy  \eqn{ufhju1}  becomes
\be
F(\bx, \bw) = \sum_{i=1}^N \left( T x_i w_i -{ c \over 2}x_i^2 - B_i x_i \right) - T \lambda(\bf w)\ .
\label{S}
\ee
The free energy is a function of both sets of variables $\bx$ and $\bw$. We can define the free energy in the $x_i$
by eliminating the $w_i$ using their saddle point equation.
This gives
\be
F(\bx) = \sum_{i=1}^N \left( T x_i w_i (\bx) -{ c \over 2}x_i^2 -  B_i x_i \right) -T \lambda(\bw(\bx))\ ,
\label{Sx}\ee
where $w_i (\bf x)$ is the solution of the $z_i$ (or $w_i$) saddle point equation in \eqn{eqsfxz}, that is,
\be
x_i  = {\del \lambda \over \del w_i}\ .
\label{w}\ee
Summing \eqn{w} over $i$ and using the first eq.\ in \eqn{chidel},
the right hand side gives the degree of homogeneity $n_\chi$ of $\chi$.
Thus we obtain 
\be
\sum_{i=1}^N x_i = \sum_{i=1}^N  {\del \lambda \over \del w_i} = n_\chi\ .
\label{sx}
\ee
This is a constraint on the variables $x_i$, consistent with \eqn{kx} and \eqn{lengyt}, necessary for \eqn{w} to have a solution for
the $w_i$. If it is
satisfied, the $w_i$ are only determined up to a common additive function. That is, the transformation 
\be
w_i (\bx) \to w_i (\bx) + f(\bx)\ ,
\label{gw}
\ee 
induces the transformation $\chi \to e^{n_\chi f}\chi$ on the character of the representation,
leaving the effective action invariant. The action has a "gauge freedom," the condition \eqn{sx} being the
corresponding Gauss law.

\subsection{General stability analysis }

The first derivatives of $F(\bx)$ in \eqn{Sx} are (we omit the $\bx$ dependence of $w_i$)
\be
{\del F \over \del x_i} =T w_i - c x_i - B_i + T \sum_j \Big( x_j -{\del \lambda \over \del w_j}\Big)\ 
{\del w_j \over \del x_i}\ .
\ee
The terms in parenthesis are zero, since $w_i$ satisfies its equilibrium equation \eqn{w} on the full
manifold of the $x_i$ (that is, off-shell for $x_i$), and can be set to zero even when taking further
$\bx$-derivatives.
The Hessian then is obtained as
\be
{\del^2 F \over \del x_i \del x_j} = T\, {\del w_i \over \del x_j} - c \ \delta_{ij}\ ,
\ee
which still depends on the magnetic fields through the equations of motion \eqn{eqsfxz}.
This is not the complete stability matrix, however, as we must also account for the constraint \eqn{sx},
whose variation is
\be
\sum_{i=1}^N \delta x_i =0\ .
\ee
To implement it, we introduce the projector
\be
\label{pij}
P_{ij} = \delta_{ij} - {1\over N}\ 
\ee
and  define the variables $\tilde x_i$ through
\be
x_i =\sum_{j=1}^N P_{ij} \tilde x_j + {n_\chi \ov N}\ .
\ee
Clearly, the $\tilde x_i$ and their variation $\d \tilde x_i$ are unconstrained.
Then, differentiating \eqn{w} with respect to $\tilde x_j$ yields
\be
\label{shpoly}
 \sum_{k,\ell=1}^N {\del^2 \lambda \over \del w_i \del w_k} {\del w_k \over \del x_\ell}P_{\ell j} =P_{ij}  \ .
\ee
Defining the matrices  
\be
\label{wijlij}
W_{ij} = {\del w_i \over \del x_j}\ ,\qq  \Lambda_{ij} = {\del^2 \lambda \over \del w_i \del w_j}\ ,
\ee
we note that $\L$ is symmetric and $P \Lambda = \Lambda P = \Lambda$,
and that \eqn{shpoly} can be expressed in matrix form as
\be
P = \Lambda W P\ .
\label{PLW}
\ee
Also,
\be
\label{Lij}
\sum_{i=1}^N \L_{ij} = \sum_{i=1}^N {\del^2 \lambda \over \del w_i \del w_j} = {\del \over \del w_j} \sum_{i=1}^N {\del \lambda \over \del w_i}
= {\del \over \del w_j} n_\chi = 0\ .
\ee
Hence, $\L$ has a zero mode, corresponding to the eigenvector $u=(1,1,\dots , 1)$, which 
is also an eigenvector of $P$ with zero eigenvalue.
The second variation of the free energy for the unconstrained variables $\tilde x_i$ defines the stability matrix as 
\be
\d^2 F = \sum_{i,j=1}^N H_{ij} \d \tilde x_i   \d \tilde x_j \ ,
\ee
where the Hessian  is
\be
H = T\, P W P - c P\ .
\ee
It always has an irrelevant zero mode, corresponding to the eigenvector $u=(1,1,\dots , 1)$.
However,  its remaining eigenvalues must
be positive, that is, $H$ should be a semi-positive matrix. This means that the eigenvalues of $W$ in the projected
space $\om _n$ must satisfy
\be
T \om_n - c \geqslant 0\ .
\label{st}
\ee
Due to \eqn{PLW} and the fact that $P$ and $\L$ commute,
 the eigenvalues of $\Lambda$ in the projected space $\lambda_n$ and the $w_n$ are related as
\be
\lambda _n \om_n = 1\ .
\ee
This implies the relation for stability
\be
0 \leqslant  \lambda_n \leqslant {T\over c}\ ,
\label{sta}
\ee
equivalently, 
\be
\Lambda \geqslant  0 \ ,\qq T\, \mathbb{1} - c \Lambda \geqslant  0\ .
\label{TL}
\ee
That is, both these matrices must be semi-positive definite.
Note that, even though the condition $\Lambda \geqslant  0$ is in principle nontrivial, it turns out that it is always identically
satisfied in the cases of interest of the present paper.


The above conditions will be analyzed in detail for $\chi$ being the fundamental, symmetric, or antisymmetric irrep.
As we shall see, $\Lambda$ and $T\, \mathbb{1} - c\Lambda$ in these cases reduce to symmetric matrices parametrized
in terms of one or two vectors defined in an auxiliary space of Euclidean, Minkowski, or generalized Minkowski signature.
The general positivity conditions for such matrices are derived in appendix \ref{appendix}.
Note also that the matrix elements $\Lambda_{ij}$ can be written in terms of the $x_i$ by using
the equations of motion \eqn{eqsfxz} as
\be
\Lambda_{ij} = \left.{\del^2 \lambda \over \del w_i \del w_j}\right|_{{\bf w} = {(\ssb c {\bf x} +{\bf B}\ssb)\ssb/\hb T}}
= \left.{1\over \chi}{\del^2 \chi \over \del w_i \del w_j}\right|_{{\bf w} = {(\ssb c {\bf x} +{\bf B}\ssb)\ssb/\hb T}}
- x_i \, x_j\ .
\label{Lx}\ee

We conclude by mentioning that the $x_i$ of the stable solutions of the equilibrium equations \eqn{eqsfxz}
determine the irrep of the phases of the ferrromagnet. For the specific atom irreps $\chi$ that will be examined
in this paper, these phases will turn out to be:
\begin{itemize} 
\item \vskip -0.3cm the {\bf singlet}, corresponding to unbroken $SU(N)$ symmetry
\item \vskip -0.2cm {\bf one-row} irreps, that is, fully symmetric irreps with a single row in their Young tableau, e.g.,
\begin{tikzpicture}[scale=.4]
\draw[thick] (0,0)--(6,0);
\draw[thick] (0,-1)--(6,-1);
\draw[thick] (0,0)--(0,-1);\draw[thick] (1,0)--(1,-1);\draw[thick] (2,0)--(2,-1);\draw[thick] (3,0)--(3,-1);
\draw[thick] (4,0)--(4,-1);\draw[thick] (5,0)--(5,-1);\draw[thick] (6,0)--(6,-1);
\end{tikzpicture} ,
breaking $SU(N)$ to $SU(N-1)\times U(1)$
\item \vskip -0.2cm
{\bf two-row} irreps, that is, doubly symmetric irreps with two {\it equal} rows in their Young tableau, e.g.,
\hskip 2.2cm ,
breaking $SU(N)$ to $SU(N-2)\times SU(2)\times U(1)$
\vskip -0.8cm
\hskip 3.75cm\begin{tikzpicture}[scale=.4]
\draw[thick] (0,-0.1)--(5,-0.10);
\draw[thick] (0,-1.1)--(5,-1.1);
\draw[thick] (0,-2.1)--(5,-2.1);
\draw[thick] (0,-0.1)--(0,-2.1);
\draw[thick] (1,-0.1)--(1,-2.1);
\draw[thick] (2,-0.1)--(2,-2.1);
\draw[thick] (3,-0.1)--(3,-2.1);
\draw[thick] (4,-0.1)--(4,-2.1);\draw[thick] (5,-0.1)--(5,-2.1);
\end{tikzpicture} 
\end{itemize}

\section{The fundamental representation: A review}\label{fundamental}

It is instructive to apply the above formalism to the known case of the fundamental representation and reproduce the 
results of \cite{PSferro}. In this case the character is
\be
\label{fjchi}
\chi({\bf z})= \sum_{i=1}^N  z_i \ ,
\ee
corresponding to a one-box YT  \hskip -.19 cm 
{\tiny \begin{ytableau}
\none  & 
\end{ytableau}} , i.e. $n_\chi =1$, resulting to the constraint 
\be
\label{hfttt}
\sum_{i=1}^N x_i = 1\ ,
\ee
in agreement with the general expression \eqn{sx}.
From the first of \eqn{eqsfxz} we get that 
\be
\label{zixi}
z_i = x_i \, \chi({\bf z})\ .
\ee
Then, the second of \eqn{eqsfxz} gives 
\be
\label{eqfund}
T \ln x_i - c x_i = B_i -T \l \ ,
\ee
where, from \eqn{lax} and \eqn{fjchi},
\be
\lambda = \ln \sum_{i=1}^N z_i = \ln\sum_{i=1}^N e^{w_i}\ .
\ee
In the equations \eqn{eqfund}, $\lambda$ can be considered a Lagrange multiplier fixed by \eqn{hfttt}.

\no
In the rest of the paper we set the magnetic fields $B_i$ to zero. 
Hence, all $x_i$ obey the same equation for fixed $\lambda$, which has two solutions $x_1$ and $x_2$
(not to be confused with $x_i$ for $i=1,2$).
Letting $p_1$ and  $p_2$ be
the number of times that each of them appears, we have
\be
\label{p1p2}
p_1+p_2=N\ ,\qq p_1 x_1 +p_2 x_2 =1\ ,
\ee
the latter due to the constraint \eqn{hfttt}.
These are solved by 
\be
\label{x1x2}
x_1 = {1+x\ov N} \ ,\qq x_2 = {1-a x \ov N}\  ,\qq a= {p_1\ov p_2} = {p_1\ov N-p_1}\ .
\ee
Clearly, $x=0$ corresponds to the singlet. Introducing a temperature scale $T_0$ as
\be
\label{cnt}
c= N T_0\ ,
\ee\uu\uu\no
we find from \eqn{eqfund} that
\be
\label{jfhj9}
\ln {1+x\ov 1-a x} - { T_0\ov T} (1+a) x= 0\ .
\ee
Note that the free energy \eqn{Sx} after using \eqn{zixi}  becomes
\be
\label{freefund}
\begin{split}
& F_{\rm fund}    = \sum_{i=1}^N T x_i \ln x_i - {c\ov 2} x_i^2
\\  \ \
&  = {T\ov 1+a}\Big(a(1+x) \ln (1+x) + (1-ax) \ln (1-ax)\Big) - {a T_0\ov 2} x^2 -T \ln N- {T_0\ov 2}  \ ,
\end{split}
\ee
where in the last step we used \eqn{x1x2}. Varying this in $x$ reproduces \eqn{jfhj9}.

\subsection{Conditions for stability }

\no
For the case of the fundamental representation the equation for $w_i$ \eqn{w} is
\be
x_i = {e^{w_i} \over \chi}
\ee
and \eqn{wijlij}, equivalently \eqn{Lx}, gives
\be
\Lambda_{ij} = {e^{w_i}\over \chi} \delta_{ij} - {e^{w_i + w_j} \over \chi^2}
= x_i \delta_{ij} - x_i x_j\ .
\label{LF}
\ee
It can be easily seen that $\Lambda \geqslant 0$ in  \eqn{TL} holds identically, as advertised. 
For the condition $T\, \mathbb{1}-c\Lambda \geqslant  0$ in \eqn{TL}, we first identify the matrix elements 
in the form \eqn{j9iug}. We have
\be
f_i = T- c x_i\ ,\qq a_i =\sqrt{c}\, x_i\ ,\qq  \e=1\ . 
\ee
According to the general stability results of appendix \ref{appendix}, this can be positive definite only if at
most one of the $f_i$ is negative.

\no
If $T-c x_i >0$ for all $i$, this corresponds the Euclidean case $s=0$ and, according to \eqn{s0s}, stability is guaranteed. 
\no
The other possibility is $T-c x_i <0$ for one of the $x_i$, say $x_1$, so that $p_1=1$,
and positive for the remaining $i$, and corresponds to the Minkowski case $s=1$. Then the condition for stability
\eqn{s1s} amounts to
\be
\sum_{i=1}^N {c x_i^2 \over T- c x_i} < -1\ ,
\ee
which, upon using \eqn{hfttt}, can be brought to the form
\be
\sum_{i=1}^N {x_i \over T- c x_i} <0\ ,
\ee
which is precisely the condition found in \cite{PSferro}. 

\no
For completeness, we restate the results of the stability analysis.
Depending on the temperature, the only stable cases correspond to the singlet and one-row representations. 
The situation is summarized in table \ref{table:0}.
\begin{table}[!ht]
\begin{center}
\begin{tabular}{|c|c|c|c|c|} \hline
  irrep & $T<T_0$ & $T_0<T<T_1 $ & $T_1<T<T_c$ & $ T_c<T$
  \\ \hline \hline
{\rm singlet}  &  {\rm unstable}      &  {\rm metastable}   &  {\rm stable }  & {\rm stable}     
\\ \hline
 {\rm one-row} & {\rm stable}    &  {\rm stable} &   {\rm metastable}    &  {\rm not a solution}     
 \\ \hline
\end{tabular}
\end{center}
\vskip -.3 cm
\caption{\small{Composing fundamental reps: Phases in various temperature ranges for $N\geqslant 3$ and their stability characterization.}}
\label{table:0}
\end{table}
The temperature $T_0$ in \eqn{cnt} was defined such that it be the temperature below which the singlet is unstable.
Similar definitions will be adopted later in the paper for the symmetric and antisymmetric irreps. The temperature $T_c$, above which
only the singlet solution exists, can be  computed by numerically solving a system of transcendental equations consisting of \eqn{jfhj9} and 
its derivative with respect to $x$ \cite{PSferro}. Between $T_0$ and $T_c$
both the singlet and one-row irreps are locally stable, and there is an intermediate temperature $T_1$ at which the
stability-metastability properties of the two solutions are exchanged. Interestingly, $T_1$ can be
analytically determined by equating the free energies for the two irreps, as 
\be
\label{tempt1}
T_1 = {T_0\ov 2} {N(N-2)\ov (N-1) \ln(N-1)}\  
\ee
and corresponds to the value $x=N-2$ for the one-row solution.  The length of the row of the YT is  
 \be
  \ell_1 \simeq (x_1-x_2)n   = {x\ov N-1} n\ .
  \ee
The spontaneous symmetry breaking in the one-row phase is
\be
SU(N)\quad {\longrightarrow} \quad SU(N-1)\times U(1)\ .
\ee
For $N=2$, it is evident that $T_1=T_0$. It turns out that the transition at $T=T_0$ is second-order, as the free energy exhibits a discontinuous second derivative with respect to $T$  \cite{PSferro}.

\no
The behavior of the ferromagnet for $N \gg 1$ was obtained in  \cite{PSferro} to be 
\be
T_c \simeq T_0 {N\ov \ln N}\ ,\qq  T_{1} \simeq {T_c \ov 2}\ .
\label{largeNSfun}
\ee

  \section{The symmetric representation}\label{symmetric}
  
\begin{figure} [th!] 
\begin{center}
\includegraphics[height= 5 cm, angle=0]{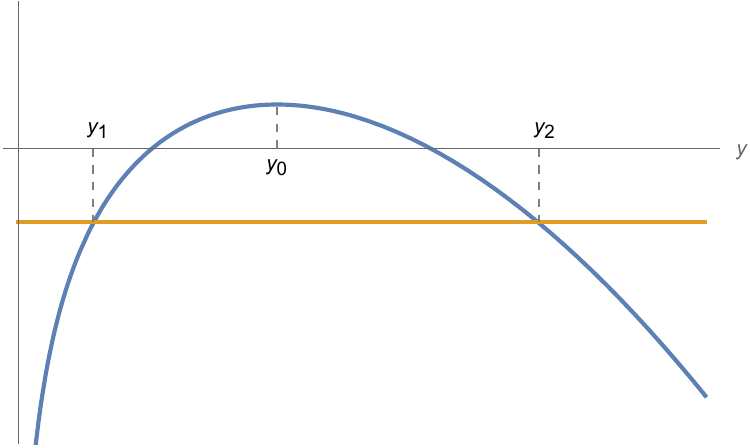}
\end{center}
\vskip -.5 cm
\caption{\small{Plots of the left hand side of \eqn{djh11} as a function of $y_i$
for generic values of $T$ and ${\bf }y^2$. The horizontal line
represent the right hand side $R = -\ln \sum_j z_j$. There are generically two solutions, $y_1 , y_2$, on either side of the
maximum occurring at $y_0$. $R$ and ${\bf y}^2$ act as Lagrange multipliers and are fixed to satisfy the two constraints,
$p_1 y_1 + p_2 y_2 = 1$ and $p_1 y_1^2 + p_2 y_2^2 = {\bf y}^2$.}}
\label{symmcurve}
\end{figure}

The usefulness of the general formalism becomes apparent for irreps higher than the fundamental. In this section we will examine the symmetric representation, for which the YT is \hskip -.19 cm {\tiny  \begin{ytableau}\none && \end{ytableau}} .
The character is \cite{PScompo}
 \be
 \label{hjks}
 \chi({\bf z}) = \ha \left(\sum_{i=1}^N z_i \right)^2 + \ha \sum z_i^2\ .
 \ee
Defining rescaled variables
\be
\label{zi}
y_i = {z_i \over \sum_{j=1}^N z_j} \ ,\qquad \sum_{i=1}^N y_i =1 \ ,
\ee
the first of \eqn{eqsfxz} expressed in the variables $y_i$ becomes
\be
\label{xiyi}
x_i = {2 \ov 1+{\bf y}^2 }\,  y_i (1+y_i)\ ,
\ee
from which, summing over $x_i$, we obtain the constraint
\be
\label{sumxi2}
\sum_{i=1}^N x_i = 2 \ ,
\ee
in agreement with \eqn{sx} since $n_{_{\scalebox{.6}{\hskip -.3 cm{\tiny \begin{ytableau}\none && \end{ytableau}}}}}= 2$.
Finally,  the second of \eqn{eqsfxz} (with $B_i=0$) gives
\be
\label{djh11}
\ln y_i - {2 c\ov T(1+{\bf y}^2)}\, y_i(1+y_i)=  -\ln \sum_{j=1}^N z_j\ .
\ee
\noindent
So, all $y_i$ obey the same equation. It can easily be seen that, for fixed right-hand side,
there are generically two solutions to this equation (fig. \ref{symmcurve}), say $y_1$ and $y_2$ 
(not to be confused with $y_i$, introduced in \eqn{zi}, for $i=1,2$). 
As in the case of a fundamental $\chi$, let $p_1$ and $p_2$ be the times each of these solutions appear,
so that 
\be
\label{jkj111}
p_1+p_2=N \ ,\qq  p_1 y_1 + p_2 y_2 = 1\ .
\ee 
To solve the above constraints, we may set 
\be
\label{jsdkn33}
y_1 = {1+y\ov N}\ ,\qq y_2 = {1-a y\ov N} \ ,\qq a= {p_1\ov p_2}={p_1\ov N-p_1}\ ,
\ee
so that 
\be
\label{jsdknss}
{\bf y}^2 = p_1 y_1^2 + p_2 y_2^2 = {1+a y^2\ov N}\ 
\ee
and define a new temperature scale $T^{(s)}_0$ by 
 \be
 \label{csymm}
 c= {T^{(s)}_0\ov 2} N {N+1\ov N+2} \ .
 \ee
We will see that $T^{(s)}_0$ is the temperature below which the singlet solution becomes unstable. 
Elimination of the right-hand side in \eqn{djh11} gives the transcendental equation
\be
\label{trans1}
\ln{1+y\ov 1-a y} - {T_0^{(s)}\ov T} (1+a) {N+1\ov N+2} \cdot{N+2+(1-a)y\ov N+1 + a y^2} y = 0\ .
\ee
Clearly the space of solutions is invariant under  the symmetry 
\be
\label{symaa}
a\to 1/a\ \ ({\rm or}\ p_1\leftrightarrow p_2) \ ,\quad  y\to -a y \ ,
\ee
and thus we may restrict to solutions with 
\be
\label{akmeros}
p_1=0,1,2,\dots ,\Big[{N\ov 2}\Big]\ ,
\ee
where the brackets stand for the integer part.
This solution corresponds to a YT with $p_1$ equal rows of length
\be
\begin{split}
\ell_1  =\dots =  \ell_{p_1} & \simeq ( x_1-x_2)n= {2\ov 1+{\bf y}^2} (y_1-y_2)(1+y_1+y_2)  n 
\\
 & = 2{1+a\ov N}{y\ov 1+N+a y^2} \big(N+2+(1-a)y\big) n\ ,
\end{split}
\ee
with $a$ as in \eqn{jsdkn33}. The above covers the singlet case with $y=0$, which obviously solves \eqn{trans1}.
Stability analysis, to which we turn next, reveals that only the singlet and the one-row cases can be stable, each for a specific temperature range. 

\subsection{Conditions for stability }

For the case of the symmetric representation, from \eqn{hjks} and \eqn{wijlij} we obtain
\be
\Lambda_{ij} = {2\ov 1+{\bf y}^2} \bigg({y_i(1+2y_i) } \d_{ij} +  {y_i y_j } -{2y_i (1+y_i) y_j (1+y_j)\ov 1+{\bf y}^2} \bigg)
\equiv {2\,{\tilde \Lambda}_{ij} \ov 1+{\bf y}^2 }\ .
\label{Ls}
\ee
As already stated, it turns out that the condition $\Lambda >0$ in \eqn{TL} is identically satisfied.
The remaining stability condition is
\be
T - c {\Lambda} \geqslant  0 ~,\quad \text{or}\quad  {T \over 2c}(1+{\bf y}^2 ) - {\tilde{\Lambda}} \geqslant 0\ .
\label{stasL}\ee
From \eqn{Ls}, the matrix elements for the second condition in \eqn{stasL} correspond to  the matrix \eqn{mij} 
with 
\be
f_i = {T \over 2c}(1+{\bf y}^2 ) - y_i (1 + 2 y_i ) \ , ~~ a_i = y_i  \ ,
~~  b_i = {\sqrt 2\, y_i (1+ y_i ) \ov\sqrt{ 1+{\bf y}^2}}\ ,~~\e_1= -1\ ,~ \e_2= 1\, .
\ee
Using this and the fact that there are two distinct solutions $y_1$ and $y_2$ we compute
\be
\begin{split}
& {\bf a}^2 = p_1 {y_1^2\ov f_1} + p_2 {y_2^2\ov f_2}\ ,
\\
&
{\bf b}^2 =   {2 \ov 1+ {\bf y}^2 } \bigg( p_1{y_1^2(1+y_1)^2\ov f_1 }
+ p_2{y_2^2(1+y_2)^2\ov f_2}\bigg)
 \ ,
\\
&
{\bf a}\cdot {\bf b}  =   \sqrt{2 \ov 1+ {\bf y}^2}\,
\bigg(p_1{y_1^2(1+y_1)\ov f_1} +  p_2{y_2^2(1+y_2)\ov f_2} \bigg) \ ,
\end{split}\label{dii}
\ee
where, using \eqn{csymm},
\be
 f_a =  {T\ov  T^{(s)}_0 } {N+2\ov (N+1)N} (1+{\bf y}^2) - y_a (1+2y_a)\ ,\qq  \ a=1,2\ .
 \ee
 In the above, $y_1$, $y_2$, and ${\bf y}^2$ should be expressed in terms of $y$ using \eqn{jsdkn33}.
 Hence, the full matrix depends on $y$, which satisfies \eqn{trans1}.
Note that, since $y_1 > y_2$ in \eqn{jsdkn33}, $f_1<f_2$.  
According to the results of appendix \ref{appendix}, since the stability matrix \eqn{mij} is parametrized by two vectors, and $\e_1=-1$ 
and $\e_2=1$, stability requires that either \eqn{s0s11}  or \eqn{jfhjh44c} hold.
Overall, the stability criteria become
\be
\begin{split}
&f_1>0\,  ,\quad  {\rm any}\ p_1\ \  \&\ \ p_2=N-p_1\, ,\quad   	\Delta  >0\ ,
\\
&f_1<0\, ,\  f_2>0\, , \quad   p_1=1 \ \  \&  \  \ p_2=N-1 \, ,\quad {\bf a}^2<1\, , \quad \Delta  >0\ ,
\label{stabcrit}\\
& \text{where}\quad \Delta = f_1 f_2 \Big((1- {\bf a}^2 )(1+  {\bf b}^2)+ ({\bf a}\cdot{\bf b})^2\Big)\ ,
\end{split}
\ee
where we note that the factor $f_2$ in $\D$ could be omitted since it is positive.

 \subsection{Solutions and their stability}
 
The value $y=0$, corresponding to the singlet state and no magnetization, is always a  solution of  \eqn{trans1} for any temperature $T$. 
For this value, $y_i=1/N$ and
\be
f_1= f_2 = {N+2\ov N^2}\bigg({T\ov {T_0^{(s)}}}-1\bigg)\ ,\qq \Delta = \left({N+2\ov N^2}\right)^{\hskip -0.1cm 2} {T\ov {T_0^{(s)}}} 
\bigg({T\ov {T_0^{(s)}}}-1\bigg)\ .
\ee
This solution is locally stable for $T>T_0^{(s)}$ and unstable for $T<T_0^{(s)}$.

\noindent
For high enough temperature, $y=0$ is the only solution of \eqn{trans1}. However, there is a temperature $T_c$
at which spontaneous magnetization occurs. The critical temperature $T_c$ is found by satisfying \eqn{trans1} and,
in addition, setting its first derivative with respect to $y$ to zero. Solving the latter equation for $T_c$ we obtain
\be
T_c = T_0^{(s)}(1 + N) (N-1- y_c) (y_c+1) {N^3 + 2 N^2 (1+y_c) - (N+2) (1+y_c)^2 \ov (N+2) (N^2 -1+y_c^2)^2}
\label{Tc}\ee
and substituting this back into  \eqn{trans1} we find the transcendental equation for $y_c$
\be
\label{trans1c}
\begin{split}
&\hskip -1.5cm{ \ln 
{(N-1)(1+y_c)\ov N-1-y_c}
 =  {N y_c (N^2-1 + y_c^2)\ov (N-1) (N-1 - y_c) (1 + y_c)}}\\
&\qq \quad\quad\quad \times {N^2 +(N-2)(1+ y_c)\ov N^3+ 2N^2(1+ y_c)-(N+2)(1+ y_c)^2} \ .
\end{split}
\ee
The critical $y_c$ and the ratio $T_c/T_0^{(s)}$ depend solely on the rank of the group $N$.

\no
For $T_0^{(s)} <T < T_c$, two nontrivial solutions of \eqn{trans1} with $y> 0$ exist. For these solutions,
$f_1 <0$ and $f_2 >0$. Therefore, according to the stability criteria \eqn{stabcrit},
stable solutions must have $p_1=1$. Similarly, for $T<T_0^{(s)}$, one nontrivial solution with $y>0$ exist,
with $f_1 <0$ and $f_2 >0$, for which we must have also $p_1 = 1$,
and one solution with $y<0$ exists, with $f_1 >0$ and $f_2 <0$, for which we must have
$p_1 = N-1$. The solution with $y<0$, however, corresponds to $y_1 <y_2$ and, due to the symmetry  \eqn{symaa}, is equivalent
to the one with $y>0$, $p_1 =1$. Therefore, we can consider only the case $p_1=1$, for which the following picture emerges
 (see fig \ref{figsta}):
 \begin{figure} [th!] 
\begin{center}
\includegraphics[height= 11 cm, angle=0]{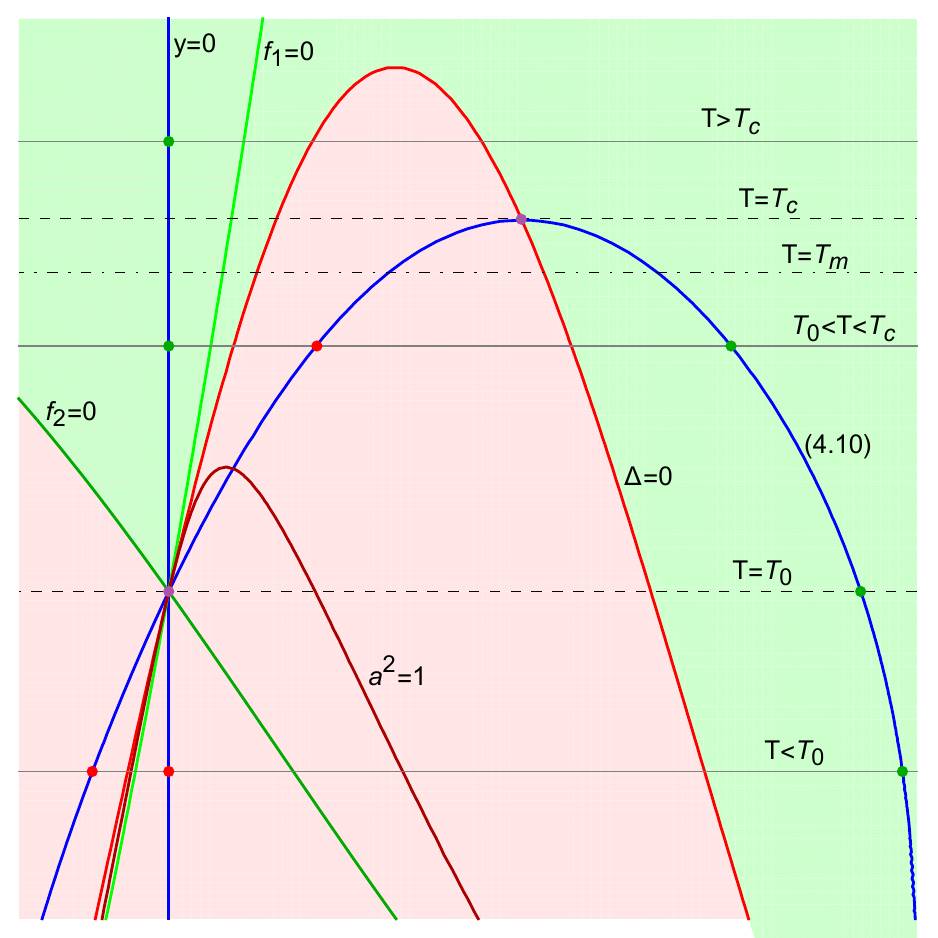}
\end{center}
\vskip -.5 cm
\caption{\small{Plot of the configurations of the symmetric-irrep ferromagnet for $N=5$, on the $y - T$ plane. The blue curve,
as well as the vertical line $y=0$, represent the solutions of \eqn{trans1}; the red curve
represents points at which the determinant condition $\Delta$ in \eqn{stabcrit} vanishes; the dark red curve represents points at which
${\bf a}^2$, appearing in \eqn{stabcrit}, becomes 1; and the light (resp. dark) green curves represent points at which
$f_1$ (resp. $f_2$) in \eqn{stabcrit} vanish. The functions $\Delta,f_1,f_2$ and $1\bb-\bb{\bf a}^2$ become positive above their
respective vanishing curves. The green shaded area represents points at which the stability criteria in
\eqn{stabcrit} are satisfied, while the red shaded area represents unstable points at which \eqn{stabcrit} fails.
Horizontal lines correspond to various temperatures, and their intersections with the curve \eqn{trans1} and $y=0$ represent solutions.
The critical temperatures $T=T_c$ and $T=T_0^{(s)}$ are marked as dashed lines, while the metastability transition
temperature $T_m$ is marked as a dotted line.
Stable solutions are marked as green dots and unstable ones as red dots. For $T=T_c$ there is a marginally stable
solution at $y=y_c$, and for $T\bb=\bb T_0^{(s)}$ there is a marginally stable solution at $y = 0$, both represented by purple dots.}}
\label{figsta}
\end{figure}
 
 \no
For $T>T_c$ there is only the solution $y=0$ and is stable.
 
 \no
For $T_0^{(s)}< T< T_c$, \eqn{trans1} has three solutions: $y=0$, as well as $y'<y_c$ and $y" >y_c$. Both $y=0$ and $y=y"$
are locally stable, one of them being globally stable and the other one metastable. The transition happens at a
temperature $T_{\rm m}$ to be computed and discussed below.

\no
For $T<T_0^{(s)}$ there are again three solutions, $y''>0$ being the only stable one.

\no
We conclude by pointing out that the stability conditions are equivalent to the sign of the $y$-derivative of the left hand side of the
equation for $y$ \eqn{trans1} being positive. This can be checked by noticing that the condition $\D=0$ leads to the same equation
for $T$ as the vanishing of the above derivative, and it parallels an equivalent relation in the fundamental irrep case. In particular,
this means that the magnetized ($y>0$) solution appearing at $T=T_c$ is marginally stable, becoming increasingly stable as
$T$ decreases. This is also clear from figure \ref{figsta}.

\subsubsection{Metastability}

The solutions that are locally stable can be either globally stable or metastable, upon comparing their free energy.
Using \eqn{hjk18} and \eqn{xiyi} in the general expression \eqn{Sx}, the free energy becomes (for $p_1 =1$)
\be
\label{freesymm}
\begin{split}
&F_{\rm sym} (T,N,p_1)   = \sum_{i=1}^N 2 T\, {y_i(1+y_i)\ov 1 +{\bf y}^2} \ln y_i - 2c\, {y_i^2(1+y_i)^2\ov (1 +{\bf y}^2)^2} - T\ln{1 +{\bf y}^2\ov 2}
\\ 
&  \qq= {2 T\ov 1 +{\bf y}^2} \Big( y_1 (1+y_1)\ln y_1 + (N-1) y_2 (1+y_2)\ln y_2\Big) 
\\
& \qq - T_0^{(s)} N {N+1\ov N+2}\, {y_1^2 (1+y_1)^2 + (N-1) y_2^2 (1+y_2)^2\ov (1 +{\bf y}^2)^2} - T\ln{1 +{\bf y}^2\ov 2}  \ ,
\end{split}
\ee
where in the second line we expressed the $y_i$ in terms of the two solutions $y_1$ and $y_2$. 
Substituting \eqn{jsdkn33} and \eqn{jsdknss} leads to a complicated expression in $y$ that we will not reproduce here.
Nevertheless, we have checked that, varying this free energy with respect to $y$ indeed reproduces \eqn{trans1} as it should.\\
For the singlet solution $y_1=y_2={\bf y}^2 = 1/N$, we obtain
\be
\label{freesinglet}
\begin{split}
 F_{\rm singlet}(T,N) = -T \ln {N(N+1)\ov 2} - T_0^{(s)} {N+1\ov N+2} \ .
\end{split}
\ee
The temperature $T_{\rm m}$ at which metastability
changes is obtained by equating
\be
 F_{\rm sym} (T_{\rm m} ,N) =   F_{\rm singlet}(T_{\rm m},N) \ .
\ee
This appears to be a hard equation to solve, due to the complicated form of $F_{\rm sym}(T_m,N)$,
but nevertheless its explicit solution can be found. 
The variable $y>0$ for the solution at $T=T_m$ assumes the simple form 
\be
y_{\rm m} = {(N+1)(N-2)\ov N+2}\ .
\label{ym}\ee
Then from \eqn{trans1} the temperature is found to be
\be
T_{\rm m}=T_0^{(s)}  {N(N+1)(N-2)\ov  (N-1)(N+2)\ln {N(N-1)\ov 2}}\ ,\qq N\geqslant 2\ .
\label{tm}\ee
The length of the single-row YT at this temperature is found by 
\be
T= T_{\rm m}:\quad x_1 =2 {N-1\ov N}\ ,\quad x_2 = {2\ov N(N-1)}\quad \Longrightarrow \quad  x_1-x_2=2{N-2\ov N-1}\ .
\ee
Hence, the length of the YT at this transition temperature is $\displaystyle 2 {N-2\ov N-1}\, n$.
Note that, for $N=2$, the temperature $T_{\rm m}=T_0^{(s)}$, in accordance with  \cite{PSspins},
where  the composition of general spin irreps of $SU(2)$ was considered.

\no
The situation is summarized in table \ref{table:1}.
\begin{table}[!ht]
\begin{center}
\begin{tabular}{|c|c|c|c|c|} \hline
  irrep & $T<T^{(s)}_0$ & $T^{(s)}_0<T<T_{\rm m}$ & $T_{\rm m}<T<T_c$ & $ T_c<T$
  \\ \hline \hline
{\rm singlet}  &  {\rm unstable}      &  {\rm metastable}   &  {\rm stable }  & {\rm stable}     
\\ \hline
 {\rm one-row} & {\rm stable}    &  {\rm stable} &   {\rm metastable}    &  {\rm not a solution}     
 \\ \hline
\end{tabular}
\end{center}
\vskip -.3 cm
\caption{\small{Phases in various temperature ranges for $N\geqslant 3$ and their stability characterization.}}
\label{table:1}
\end{table}
\vskip -0.5cm
\noindent
Hence, we have the symmetry breaking pattern for the globally stable configuration
\be
SU(N)\quad \stackrel{T<T_{\rm m}}{\hskip -0.1cm\hookrightarrow} \quad SU(N-1)\times U(1)\ .
\ee
Using \eqn{ym} and \eqn{tm}, the free energy \eqn{freesymm} near $T=T_{\rm m}$ has the expansion 
\be
\begin{split} 
& F_{\rm sym} =  -T_{\rm m} \ln {N(N+1)\ov 2} - T_0^{(s)} {N+1\ov N+2} + C^{\rm sym}_N (T-T_{\rm m}) +\dots \ ,
\\
&  C^{\rm sym}_N 
 = -{2\ov N}\ln {N \ov 2} +{N-2\ov N} \ln (N-1) - \ln (N+1)\ .  
\end{split}
\ee
{Comparing with \eqn{freesinglet} we see that, for $N\geqslant 3$, there is a discontinuity in the first derivative of the free energy when
transitioning between states. Thus, the transition at $T_{\rm m}$ is of first order.
This is generic for metastability transitions, in which the free energies
of the two phases cross at $T_{\rm m}$ with unequal first derivatives. Deviating from this would require {\it both}
the first and second derivatives of the free energy in each phase to match at the transition temperature, for a third-order
transition, or a nonanalyticity of either of the free energies at the transition, neither of which happens here.

\no
For $N=2$, the first derivatives match. Keeping also the quadratic term, we obtain
\be
N=2: \quad T_{\rm m} = T_0^{(s)},\quad F_{\rm sym} = -{3\ov 4} T_0^{(s)} -T \ln 3  -{\big(T-T_0^{(s)}\big)^2\ov T_0^{(s)}}+\dots 
\ee
in accordance with the fact that for $SU(2)$ there is no metastability transition but, rather, a single
second order phase transition at $T_0^{(s)}$ \cite{PSspins}.

\subsubsection{Small $T$ and large $N$ limits} 
A small temperature analysis of \eqn{trans1} gives the low-temperature behavior
\be
y = (N-1)\bigg( 1 - {N}\, e^{-{N(N+1)\ov (N+2)}\, { T_0^{(s)}/ T}} +\dots  \bigg)\ .
\ee
Then, up to exponentially small corrections, 
$x_1\simeq {2}$ and $x_2\simeq 0$, which is a symmetric irrep with a single YT row of length $2n$,
corresponding to maximal magnetization.

\no
We can also determine the behavior of the ferromagnet for $N \gg 1$. We obtain
\be
y_c \simeq N\bigg(1-{2\ov \ln N}\bigg) \ ,\qq T_c \simeq T_0^{(s)} {N\ov \ln N}\ ,\qq  T_{\rm m} \simeq {T_c \ov 2}\ .
\label{largeNS}\ee
We note that, at $N\gg 1$, $T_c$ and $T_{\rm m}$ are the same as those for the fundamental irrep ferromagnet \eqn{largeNSfun} assuming equal $T_0$'s.

\section{The antisymmetric representation}\label{antisymmetric}

We consider the ferromagnet with the irrep of atoms $\chi$ being the antisymmetric representation, for which the YT is 
\hskip -.3 cm 
{\tiny \begin{ytableau}
     \none    & \\
   \none  &\\
\end{ytableau}}\hskip 0.15cm and the character is
 \be
 \label{hjk18}
 \chi({\bf z}) = \ha \left(\sum_{i=1}^N z_i \right)^2 - \ha \sum z_i^2\ .
 \ee
Working as in the symmetric irrep, we define rescaled variables
\be
\label{zia}
y_i = {z_i \over \sum_{j=1}^N z_j}\ ,\qquad \sum_{i=1}^N y_i =1 \ .
\ee
Then the first equation in \eqn{eqsfxz} expressed in the variables $y_i$ becomes
\be
\label{xiyi}
x_i = {2 \ov 1-{\bf y}^2 }\,  y_i (1-y_i)\ ,
\ee
from which we obtain the constraint
\be
\label{sumxi2}
\sum_{i=1}^N x_i = 2 \ ,
\ee
again in agreement with \eqn{sx} since $n_{_{\scalebox{.55}{\hskip -.3 cm{\tiny\begin{ytableau}
     \none    & \\
   \none  &\\
\end{ytableau}}}}} =2$.
The second of \eqn{eqsfxz} (with $B_i=0$) gives
\be
\label{djh12}
\ln y_i - {2 c\ov T(1-{\bf y}^2)}\, y_i(1-y_i)=  -\ln \sum_{j=1}^N z_j\ ,
\ee
\noindent
so, all $y_i$ obey the same equation, as in the symmetric case. The crucial difference is that, now,
for fixed right-hand side, eq.~\eqn{djh12} can have either one or three solutions, depending on
the temperature. Specifically, for $T\bb>\bb c/(4(1-{\bf y}^2))$ there will be one solution, but for
$T\bb<\bb c/(4(1-{\bf y}^2))$ there can be either one or three solutions (fig. \ref{asymmcurve}).
This makes the analysis of the model more involved than for the fundamental or the symmetric irreps.
\begin{figure} [th!] 
\begin{center}
\includegraphics[height= 4.5 cm, angle=0]{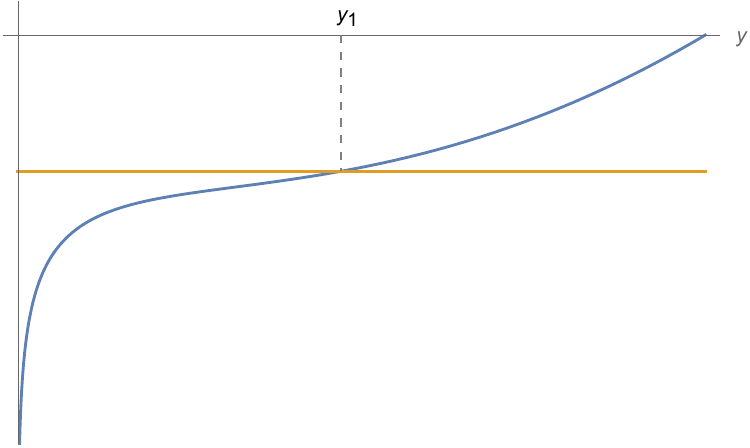} \hskip 0.6cm
\includegraphics[height= 4.5 cm, angle=0]{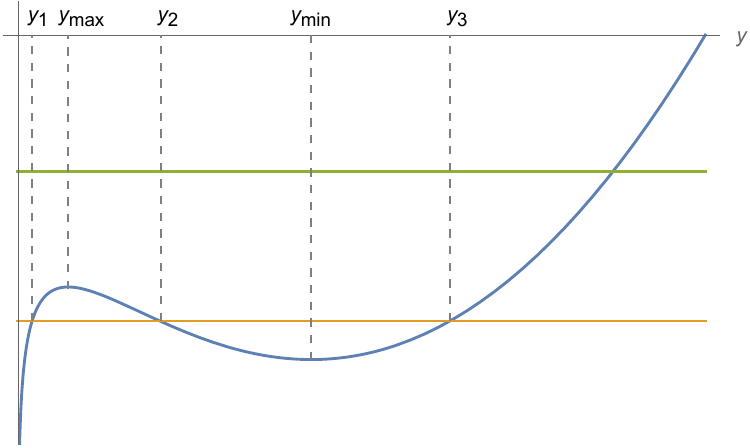}
\end{center}
\vskip -.5 cm
\caption{\small{Plots of the left hand side of \eqn{djh12} as a function of $y_i$
for $T\bb>\bb c/(4(1-{\bf y}^2))$ (left plot) and $T\bb<\bb c/(4(1-{\bf y}^2))$ (right plot). Horizontal lines
represent the right hand side $R = -\ln \sum_j z_j$. For $T\bb>\bb c/(4(1-{\bf y}^2))$ there is one solution, $y_1$.
For $T\bb<\bb c/(4(1-{\bf y}^2))$ the function develops a maximum at $y_{\rm max}$ and a minimum at $y_{\rm min}$
and we can have either one solution or three solutions
$y_1\bb<\bb y_{\rm min}\bb <\bb y_2 \bb<\bb y_{\rm max}\bb <\bb y_3$. $R$ and
${\bf y}^2$ act again as Lagrange multipliers and are fixed to satisfy the two constraints,
$p_1 y_1 + p_2 y_2 + p_3 y_3= 1$ and $p_1 y_1^2 + p_2 y_2^2 + p_3 y_3^2= {\bf y}^2$.}}
\label{asymmcurve}
\end{figure}

\no
We can generically assume that there are three solutions, say $y_1,y_2,y_3$
(again not to be confused with $y_i$ in \eqn{zia} for $i=1,2,3$), appearing $p_1, p_2, p_3$ times
respectively in \eqn{djh12},
since we can always take some of the $y_1, y_2, y_3$ to be equal, or some of the $p_1, p_2, p_3$ to vanish,
reducing the situation to the one with two or one solutions. The constraints on $y_i$ imply
\vskip-1cm\be
\label{jkj111}
p_1+p_2+p_3=N \ ,\quad  p_1 y_1 + p_2 y_2 +p_3 y_3= 1\ ,\quad p_1 y_1^2 + p_2 y_2^2 + p_3 y_3^2 = {\bf y}^2\, .
\ee
We can solve the first two equations for $p_3$ and $y_3$ and substitute in ${\bf y}^2$.
The equations \eqn{djh12}, then, for $y_1,y_2,y_3$ become three equations for the variables $y_1,y_2$ and
$\ln \sum_i z_i$. Eliminating this last quantity leads to the system of transcendental equations for the two
variables $y_1$ and $y_2$
\be
\begin{split}
& \ln{y_1 \ov y_2} = {2 c\ov T (1 - {\bf y}^2)} \Big(y_1 (1 - y_1) - y_2 (1 - y_2) \Big) \ ,
\\
& \ln{y_1 \ov y_3} = {2 c\ov T (1 - {\bf y}^2)} \Big(y_1 (1 - y_1) - y_3 (1 - y_3) \Big) \ ,
\\
& y_3 = {1-p_1 y_1 - p_2 y_2 \ov p_3}~,\quad {\bf y^2} = p_1 y_1^2 + p_2 y_2^2 + p_3 y_3^2\ .
\label{transs}\end{split}
\ee
Unlike in the symmetric case, we cannot reduce the above system to a unique
transcendental equation in one variable. For later convenience, we also define the new temperature scale $T^{(a)}_0$
 \be
 \label{cta}
 c= {T^{(a)}_0\ov 2} N {N-1\ov N-2} \ ,\qq N\geqslant 3\ ,
 \ee
which will turn out to be the temperature at which the singlet state becomes unstable.

\subsection{Conditions for stability}

For the case of the antisymmetric representation, from \eqn{hjk18} and \eqn{wijlij} we obtain
\be
\Lambda_{ij} = {2\ov 1-{\bf y}^2} \bigg({y_i(1-2y_i) } \d_{ij} +  {y_i y_j } -{2y_i (1-y_i) y_j (1-y_j)\ov 1-{\bf y}^2} \bigg)
\equiv {2\, {\tilde \Lambda}_{ij} \ov 1-{\bf y}^2 }\ .
\label{La}
\ee
As in the symmetric and fundamental irreps, the condition $\Lambda \geqslant 0$ in \eqn{TL} is identically satisfied.
The remaining stability condition is
\be
T - c {\Lambda} \geqslant  0 \ ,\quad \text{or}\quad  {T \over 2c}(1-{\bf y}^2 ) - {\tilde{\Lambda}}\geqslant  0\ .
\label{staaL}\ee
From \eqn{La}, the matrix elements for the second condition in \eqn{staaL} correspond to  the matrix \eqn{mij} 
with 
\be
f_i = {T \over 2c}(1-{\bf y}^2 ) - y_i (1 - 2 y_i ) \ , ~~ a_i = y_i  \ ,
~~  b_i = {\sqrt 2\, y_i (1- y_i ) \ov\sqrt{ 1-{\bf y}^2}}\ ,~~\e_1= -1\ ,~ \e_2= 1\, .
\ee
From this, for three solutions $y_1, y_2, y_3$ related as in \eqn{jkj111} we compute
\be
\begin{split}
& {\bf a}^2 = p_1 {y_1^2\ov f_1} + p_2 {y_2^2\ov f_2}+ p_3 {y_3^2\ov f_3},
\\
&
{\bf b}^2 =   {2 \ov 1- {\bf y}^2 } \bigg( p_1{y_1^2(1-y_1)^2\ov f_1 } +p_2{y_2^2(1-y_2)^2\ov f_2} 
+ p_3{y_3^2(1-y_3)^2\ov f_3 }\bigg) \ ,
\\
&
{\bf a}\cdot {\bf b}  =   \sqrt{2 \ov 1- {\bf y}^2}\,
\bigg(p_1{y_1^2(1-y_1)\ov f_1} +  p_2{y_2^2(1-y_2)\ov f_2} + p_3{y_3^2(1-y_3)\ov f_3} \bigg) \ ,
\end{split}\label{dii}
\ee
where, using \eqn{cta},
\be
 f_a =  {T\ov  T^{(s)}_0 } {N-2\ov N(N-1)} (1-{\bf y}^2) - y_a (1-2y_a)\ ,\qq  \ a=1,2,3\ .
 \ee
Similarly to the symmetric case, according to the results of appendix \ref{appendix}, and since the stability matrix \eqn{mij} is  parametrized by two-vectors 
and $\e_1=-1$ and $\e_2=1$, stability requires that either \eqn{s0s11} or \eqn{jfhjh44c} hold.
Assuming $f_1 \leqslant  f_2 \leqslant  f_3$ (which induces a particular order in the $y_i$'s), 
the stability criteria of \ref{appendix}  become 
\be
\begin{split}
&f_1,\ f_2, \ f_3>0\, ,\quad\quad  {\rm any}\ p_1,\ p_2, \ p_3,\ \  \ \ \quad   	\Delta  >0\ ,
\\
&f_1<0\, ,\  f_2, \ f_3>0\, , \quad   p_1=1 \ \  \&  \  \ {\rm any}\ \  p_2+ p_3=N-1\, , \quad {\bf a}^2<1\, , \quad \Delta  >0\ ,
\label{stabcrita}\\
& \text{where}\quad \Delta = f_1 f_2 f_3 \Big((1- {\bf a}^2 )(1+  {\bf b}^2)+ ({\bf a}\cdot{\bf b})^2\Big)\ ,
\end{split}
\ee
where the factor $f_2f_3$ in $\D$ could be omitted since it is positive.
The above is the general case with three potentially different solutions. However, the analytical treatment
of the transcendental system of equations \eqn{transs} plus the stability conditions \eqn{stabcrita} becomes intractable.
Therefore, we resorted to a numerical study of the problem to gain some intuition on the solutions.

\no
A numerical study of \eqn{transs} for generic values of $N$
shows that locally stable solutions with all $y_1,y_2,y_3$ distinct exist only for $p_1 = p_2 = 1$. In this case, the
solution corresponds to a YT with two unequal rows. However further study of these solutions reveals that they
are unstable. Therefore, only solutions with at most two distinct $y_1,y_2$ can be stable. We focus our
analysis on this case, which parallels the one for the fundamental and symmetric cases. Setting again
\be
y_1 = {1+y\ov N}\ ,\qq y_2 = {1-a y\ov N} \ ,\qq a= {p_1\ov p_2}={p_1\ov N-p_1}\ ,
\label{yya}
\ee
equations \eqn{transs} reduce to the single equation
\be
\label{fhu11}
\ln{1+y\ov 1-a y} - {T_0^{(a)}\ov T} (1+a) {N-1\ov N-2}\cdot {N-2-(1-a)y\ov  N-1 - a y^2} y = 0\ .
\ee
As in the symmetric case, the space of solutions enjoys the symmetry $p_1 \leftrightarrow N-p_1$, so we may restrict $p_1 \leqslant [N/2]$.
Note, further, that \eqn{fhu11} for the antisymmetric case is related to the corresponding
equation \eqn{trans1} for the symmetric case through the transformation $N \to -N$, establishing a formal
"duality" between the two cases.

\no
The stability criteria \eqn{stabcrita} in the case of only two distinct $y_1,y_2$ reduce to
\be
\begin{split}
&f_1,\  f_2>0 \, ,\quad  {\rm any}\ \ p_1\ ,\ p_2=N-p_1\, ,  \quad   	\Delta  >0\ ,
\\
&f_1<0\, ,\  f_2>0\ , \quad   p_1=1\ ,  \ p_2=N-1\, ,  \quad {\bf a}^2<1\, , \quad \Delta  >0\ ,
\label{stabcrita2}
\\
& \text{where}\quad \Delta = f_1 f_2 \Big((1- {\bf a}^2 )(1+  {\bf b}^2)+ ({\bf a}\cdot{\bf b})^2\Big)\ ,
\end{split}
\ee
with
\vskip-1cm\be
\begin{split}
& f_a =  {T\ov  T^{(a)}_0 } {N-2\ov N(N-1)} (1-{\bf y}^2) - y_a (1-2y_a)\ ,\qq  \ a=1,2\ ,
\\
& {\bf a}^2 = p_1 {y_1^2\ov f_1} + p_2 {y_2^2\ov f_2}\ ,
\\
&
{\bf b}^2 =   {2 \ov 1- {\bf y}^2 } \bigg( p_1{y_1^2(1-y_1)^2\ov f_1 } +p_2{y_2^2(1-y_2)^2\ov f_2} \bigg) \ ,
\\
&
{\bf a}\cdot {\bf b}  =   \sqrt{2 \ov 1- {\bf y}^2}\,
\bigg(p_1{y_1^2(1-y_1)\ov f_1} +  p_2{y_2^2(1-y_2)\ov f_2}\bigg) \ ,
\end{split}\label{dii2}
\ee
and $y_1,y_2$ as in \eqn{yya}.

 \subsection{Solutions and stability}
 
 \no
The value $y=0$, corresponding to the singlet and no magnetization, is a  solution of  \eqn{fhu11}  for any temperature $T$. Since, for $y=0$,
\be
f_1= f_2 = {N-2\ov N^2}\cdot{T-T_0^{(s)}\ov T_0^{(s)}}\ ,\qq \Delta = \left({N-2\ov N^2 T_0^{(a)}}\right)^2 \hskip -0.2cm
T (T-T_0^{(s)})\ ,
\ee
this solution is locally stable for $T>T_0^{(a)}$ and unstable for $T<T_0^{(a)}$.
As in the symmetric case, it remains the only solution to \eqn{fhu11} down to a critical temperature $T=T_c$ at which
spontaneous magnetization appears. Following an identical  procedure as in the symmetric case, the transcendental
equation for the corresponding value $y=y_c$ is
\be
\label{trans1cv}
\ln {1+y_c\ov 1-a y_c} =  
{(1+a) y_c\ov (1+y_c)(1-a y_c)}    {\big(N-2-(1-a) y_c\big)(N-1-a y_c^2)\ov (N-2)(N-1 + a y_c^2)- 2 (N-1)(1-a)y_c }\ ,
\ee
with $a$ as in \eqn{yya}. Substituting $y=y_c$ in \eqn{fhu11} gives the critical temperature $T_c$
\be
T_c = T_0^{(a)} (N-1)(1+y_c)(1-a y_c){(N-2)(N-1+a y_c^2)-2(N-1)(1-a)y_c \ov (N-2)(N-1-a y_c^2)^2}\ .
\ee

\no
It turns out that stable solutions occur only for  $p_1=1$ and $p_1=2$, the former leading to a
single-line YT (symmetric irrep) and the latter to a YT with two equal lines (doubly symmetric irrep).
Depending on the temperatures at which each phase is stable, we can have mixed states where any of the three
phases (including the singlet) can coexist, as well as various transitions between the phases.
We will examine the existence and stability of the $p_1=1$ and $p_1=2$ phases separately and combine them
to obtain the complete phase structure.

\no
For  $p_1=1$ and $T_0^{(a)}< T< T_c^{(1)}$ there are three solutions: the singlet for $y=0$ and
two symmetric ones for $y_1<y_c^{(1)}<y_2$, of which $y=0$ and $y=y_2$ are locally stable and $y=y_1$ is unstable 
($y_1$ and $y_2$ should not be confused with the two distinct solutions of \eqn{djh12}).
Their metastability properties are swapped at an intermediate
temperature $T_{\rm m}^{(01)}$ above which the singlet is globally stable and the symmetric one metastable.\footnote{In
general, $T_{\rm m}^{(ij)}$ will denote the transition temperature from an $i$-row metastable state to a $j$-row stable state.
For the phases of the present ferromagnet, irrep, $i,j=0,1,2$, $i=0$ being to the singlet.}
At temperatures below $T_0^{(a)}$ and above another critical temperature $T_0^{(1)}$, there are three solutions: $y=0$
as well as $y_1<0<y_2$, of which only $y=y_2$ is stable. For $T < T_0^{(1)}$, none of the $p_1=1$ solutions are stable.
Hence, the stable solution must correspond to the case $p_1=2$. The phases for $p_1=1$ are summarized
in table \ref{table:2}.
\begin{table}[!ht]
\small{
\begin{center}
\begin{tabular}{|c|c|c|c|c|c|} \hline
  irrep  &  $0\!<\!T\!<\!T^{(1)}_0$   &  $T_0^{(1)}\!<\!T\!<T^{(a)}_0$  & 
  $T^{(a)}_0\!<T\!<\!T^{(1)}_c$ & $ T^{(1)}_c\!<\!T$
  \\ \hline \hline
{\rm singlet}    &  {\rm unstable} &  {\rm unstable}      &  {\rm locally stable }  & {\rm locally stable}     
\\ \hline
 {\rm one-row}  &  {\rm unstable}  & {\rm locally stable}    &   {\rm locally stable}    &  {\rm not a solution}     
 \\ \hline
\end{tabular}
\end{center}}
\vskip -.3 cm
\caption{\small{Phases for the one-row solution with $p_1=1$  in relation to the singlet  in various temperature ranges for $N\geqslant 4$ and and their stability characterization.}}
\label{table:2}
\end{table}

\no
We also point out that the equations for $y_c$ and $T_c$ for $p_1 =1$ simplify and become
\be
\begin{split}
& \ln{(N\hm 1)(1\hp y_c)\ov N\hm 1\hm y_c} = {N y_c (N\hm 1\hp y_c)\ov (N\hm 1)(1\hp y_c)(N\hm 1\hm y_c )}\ ,
\\
& T_c = T_0^{(a)} (N\hm 1)(1\hp y_c ){N\hm 1\hm y_c \ov (N\hm 1 \hp y_c )^2}\ .
\end{split}
\ee

\no
The situation for $p_1=2$ is similar to the one for $p_1=1$, with the difference that for  $T<T_0^{(a)}$
the solution with $y=y_2$ is stable all the way down to zero temperature.
The situation is summarized in table \ref{table:3}. 
\begin{table}[!ht]
\small{
\begin{center}
\begin{tabular}{|c|c|c|c|c|} \hline
  irrep  &  $0<T<T^{(a)}_0$  & {$T^{(a)}_0<T<T^{(2)}_c$} & $ T^{(2)}_c<T$
  \\ \hline \hline
{\rm singlet} &  {\rm unstable}       &  {\rm locally stable }  & {\rm locally stable}     
\\ \hline
 {\rm two-row}   & {\rm locally stable}    &   {\rm locally stable}    &  {\rm not a solution}     
 \\ \hline
\end{tabular}
\end{center}}
\vskip -.3 cm
\caption{\small{Phases for the two equal-row solution with $p_1=2$ in relation to the singlet  in various temperature ranges for $N\geqslant 4$ and and their stability characterization.}}
\label{table:3}
\end{table}

\no
The complete picture emerges after considering transitions between stable and metastable states, to which we turn below.

\subsubsection{Metastability}

To determine which of the locally stable solutions is globally stable and which are metastable
we examine their free energy. From the general expression \eqn{Sx}, using \eqn{hjk18} and \eqn{xiyi},
 the free energy becomes (for general $p_1$ and $p_2=N-p_1$)
\be
\label{freeantis}
\begin{split}
 F_{\rm ant} (T,N,p_1) 
&  = {2 T\ov 1-{\bf y}^2} \Big( p_1 y_1 (1-y_1)\ln y_1 + p_2 y_2 (1-y_2)\ln y_2\Big) 
\\
& \quad - T_0^{(a)} N {N-1\ov N-2}\, {p_1 y_1^2 (1-y_1)^2 + p_2 y_2^2 (1-y_2)^2\ov (1 -{\bf y}^2)^2} - T\ln{1 -{\bf y}^2\ov 2}  \ ,
\end{split}
\ee
Substituting \eqn{yya} leads to a complicated expression in $y$, yielding \eqn{fhu11} after variation with respect to $y$,
 which we will not reproduce here.\\
\no
For the singlet solution, $y_1=y_2={\bf y}^2 = 1/N$, we obtain
\be
\label{freesingleta}
\begin{split}
 F_{\rm singlet}(T,N) = -T \ln {N(N-1)\ov 2} - T_0^{(a)} {N-1\ov N-2} \ .
\end{split}
\ee
Hence, the temperature $T^{(0p_1)}_{\rm m}$ is obtained by equating
\be
 F_{\rm ant} (T^{(0p_1)}_{\rm m},N,p_1) =   F_{\rm singlet}(T^{(0p_1)}_{\rm m},N) 
\ee
As in the symmetric case, this is a complicated equation due to the nontrivial form of $F_{\rm ant} (T,N,p_1)$,
but nevertheless its explicit solution can be found in the two cases for which global stability occurs, that is, for $p_1=1,2$.
The variable $y$ at the metastability transition temperatures assumes the particularly simple form
\be
y_{\rm m}^{(01)} = {(N-1)(N-4)\ov N}\ ,\qq y_{\rm m}^{(02)} = {(N-1)(N-4)\ov 2(N-2)}\ .
\ee
Then from \eqn{fhu11} the transition temperatures are found to be
\be
T^{(01)}_{\rm m}=T_0^{(a)}  {N(N-4)\ov 4 (N-2)\ln {N-2\ov 2}}\, ,\quad
T^{(02)}_{\rm m}=T_0^{(a)}  {N(N-1)(N-4)\ov 2 (N-2)^2\ln {(N-2)(N-3)\ov 2}}\, ,\quad
N\geqslant 3\ .
\label{ta012}\ee
The length of the single-row YT at temperature $T_{\rm m}^{(01)}$ is found by evaluating 
\be
T= T_{\rm m}^{(01)}:\quad x_1 ={N-2\ov N}\ ,\quad x_2 = {N+2\ov N(N-1)}\quad \Longrightarrow \quad  x_1-x_2={N-4\ov N-1}\ .
\ee
Hence, the length of the YT at this temperature is $\displaystyle {N-4\ov N-1}\, n$.
Similarly, at temperature $T_{\rm m}^{(02)}$ we evaluate 
\be
T= T_{\rm m}^{(02)}:\quad x_1 ={N-2\ov N}\ ,\quad x_2 = {4\ov N(N-2)}\quad \Longrightarrow \quad  x_1-x_2={N-4\ov N-2}\ .
\ee
Hence, the length of the two equal rows in the YT at this temperature is $\displaystyle {N-4\ov N-2}\, n$.

\no
As noted in the previous section, there must be a metastability transition between the one-row and two-row solutions,
occurring at a temperature $T^{(12)}_{\rm m} $ larger than the temperature $T_0^{(1)}$ at which the one-row solution
becomes unstable. This temperature satisfies the equation
\be
 F_{\rm ant} (T^{(12)}_{\rm m},N,1) =   F_{\rm ant}(T^{(12)}_{\rm m},N,2)\ .
\ee
Unlike in the cases for $T^{(01)}_{\rm m}$ and $T^{(02)}_{\rm m}$, $T^{(12)}_{\rm m}$ can only be 
calculated numerically.\footnote{
As an example, consider $N=5$. Numerically, in units of $T_0^{(a)}$, we find\vskip -0.4cm
\be
~~T_0^{(1)}\simeq 0.8747\, , \ \  T_{\rm m}^{(12)}\simeq   1.0012 \, ,\ \ T_{\rm m}^{(02)}\simeq 1.0114\, , 
\ \ T_c^{(2)}\simeq 1.0127\, ,\ \ T_{\rm m}^{(01)}\simeq 1.0276\, ,  \ T_c^{(1)}\simeq 1.0311\ .
\nonumber\ee
}
The complete picture is summarized  in table \ref{table:4} below.
\vskip0.4cm\begin{table}[!ht]
\small{
\begin{center}
\begin{tabular}{|c|c|c|c|c|c|} \hline
  Temperature  &  singlet  & one-row & two-row
  \\ \hline \hline
$T_c^{(1)}< T $ &  {\rm stable}      &  {\rm not a solution}   &  {\rm not a solution }  
\\ \hline
$T_{\rm m}^{(01)} < T< T_c^{(1)}$ &  {\rm stable}      &  {\rm metastable}   &  {\rm not a solution }  
\\ \hline
$T_c^{(2)} < T< T_{\rm m}^{(01)}$ &   {\rm metastable}     &  {\rm stable}  &  {\rm not a solution }  
\\ \hline
$T_{\rm m}^{(02)} < T< T_c^{(2)}$ &   {\rm metastable}${}^*$     &  {\rm stable}  &  {\rm metastable}   
\\ \hline
$T_{\rm m}^{(12)} < T< T_{\rm m}^{(02)}$ &   {\rm metastable}     &  {\rm stable}  &  {\rm metastable}${}^*$    
\\ \hline
$T_{0}^{(a)} < T< T_{\rm m}^{(12)}$ &   {\rm metastable}     &   {\rm metastable}${}^*$  &  {\rm stable} 
\\ \hline
$T_{0}^{(1)} < T< T_{0}^{(a)}$ &   {\rm unstable}    &   {\rm metastable}  &  {\rm stable} 
\\ \hline
$T< T_{0}^{(1)}$ &   {\rm unstable}    &   {\rm unstable}  &  {\rm stable} 
\\ \hline
\end{tabular}
\end{center}}
\vskip -.3 cm
\caption{\small{Phases, stable and metastable, for the singlet, one-row and the two-row solutions for
$N\geqslant 4$. If, in a given temperature range there are two metastable solutions, the one with
lower free energy is indicated with a star.}}
\label{table:4}
\end{table}

\no
Omitting metastable states, we can tabulate the stable solutions in the simpler table \ref{table:5}.
\begin{table}[!ht]
\small{
\begin{center}
\begin{tabular}{|c|c|} \hline
 Temperature &  irrep 
  \\ \hline \hline
$T_{\rm m}^{(01)} < T $ &  {\rm singlet}       
\\ \hline
$T_{\rm m}^{(12)} < T< T_{\rm m}^{(01)}$ & one-row   
\\ \hline
$ T< T_{\rm m}^{(12)}$ &   {\rm two-row}       
\\ \hline
\end{tabular}
\end{center}}
\vskip -.3 cm
\caption{\small{Phases for the singlet, one-row and the two-row solutions for $N\geqslant 4$ and their stability characterization excluding metastable phases.}}
\label{table:5}
\end{table}
\\
Hence, we have the spontaneous symmetry breaking patterns
\be
SU(N) \ \stackrel{T_{\rm m}^{(12)}< T<T_{\rm m}^{(01)}}{\longrightarrow}  SU(N-1)\times U(1)  \stackrel{~~~T<T^{(12)}_{\rm m}}{\longrightarrow} 
SU(N-2)\times SU(2)\times U(1) \ .
\ee

\no
The free energy \eqn{freeantis} near $T=T_{\rm m}^{(0p_1)}$ admits the expansion 
\be
 F_{\rm ant} =  - T_{\rm m}^{(0p_1)} \ln {N(N-1)\ov 2} - T_0^{(a)} {N-1\ov N-2} + C^{\rm ant}_{p_1,N} \big(T-T_{\rm m}^{(0p_1)}\big) + \dots \ ,
\ee
where $p_1=1$ or $p_1=2$ and 
\be
\begin{split}
&  C^{\rm ant}_{1,N} = - \ln{N(N-1)\ov 2} +{N-4\ov N} \ln{N-2\ov 2}\ ,
\\
& C^{\rm ant}_{2,N} = - \ln{N(N-1)\ov 2} +{N-4\ov N} \ln{(N-2)(N-3)\ov 2} \ .
\end{split}
\ee
Comparing with the free energy of the singlet \eqn{freesingleta}, we see that, for $N\geqslant 5$,
there is a discontinuity in the
first derivative of the free energy in the metastability transitions at $T_{\rm m}^{(01)}$ and $T_{\rm m}^{(02)}$.
A similar result is numerically obtained for the metastability transition at $T_{\rm m}^{(12)}$.
Thus, all metastability transitions are of first order, as in the symmetric irrep.

\subsubsection{The special cases $N=2,3,4$}

As a general remark, a ferromagnet with $\chi'({\bf z}) = (z_1 \cdots z_N )^m \chi ({\bf z})$ is equivalent to the one with
$\chi ({\bf z})$ , as
the difference amounts to an overall $U(1)$ charge. Their saddle point solutions are related by $x_i' = x_i + m$ and have identical thermodynamic
properties (up to an irrelevant constant in the energy). In particular, choosing $m=-n_\chi$ turns the irrep $\chi$ into its conjugate,
which is therefore thermodynamically equivalent to $\chi$.

\no
It is instructive to examine the special cases of $SU(2)$, $SU(3)$, and $SU(4)$.

\no
For $SU(2)$, the antisymmetric irrep is equivalent to the singlet, and therefore the system is trivial (it has a unique state).
Indeed, for $N=2$, the two components of $y_i$ are $y_1$ and $1-y_1$. Substituting in \eqn{djh12} and subtracting we obtain
\be
\ln {y_1 \ov 1-y_1} = 0 ~~~\Rightarrow~~~ y_1 = y_2 = \half \ ,
\ee
and from \eqn{xiyi} $x_1 \bb=\bb x_2 \bb=\bb 1$. Substituting in \eqn{eqsfxz} we find $z_1 = z_2 = e^{c/T}$ and $\chi = e^{2c/T}$.
Finally, from \eqn{ufhju1} we obtain $F = -c$. This is a temperature-independent free energy, corresponding to
a single state of energy $-c$. In this case \eqn{cta} gives $T_0^{(a)} = 0$.

\no
For $SU(3)$, the antisymmetric irrep is equivalent to the antifundamental, so its thermodynamics should be the same
as the fundamental. First, note that \eqn{cnt} and \eqn{cta} imply $T_0^{(a)} = T_0$. 
For $N=3$ the only nontrivial case is $p_1=1$, so $a=1/2$, and 
letting $y=-{2x\ov 2+x}$, eq.
\eqn{fhu11} 
becomes the transcendental equation \eqn{jfhj9}
with $a=1/2$. Further, \eqn{yya}, \eqn{xiyi} and
\eqn{x1x2} identify $x=-{2y \ov 2+y}$ as the corresponding parameter of the fundamental irrrep with the same free energy.
As expected, $T^{(01)}_{\rm m}$ reproduces the expression \eqn{tempt1}  for the fundamental of $SU(3)$.

\no
The case of $SU(4)$ is more interesting: the antisymmetric irrep is equivalent to its conjugate,
and this has special implications for the configurations of the model.
One-row representations, which appear as states of the model for generic $N$, are conjugate to representations with three
equal rows, which are not stable solutions for the antisymmetric irrep. Therefore, the one-row solution must be absent for $N=4$,
leaving only the singlet and two-row representations as possible states.
Explicit calculation shows that all critical temperatures collapse to $T_0^{(a)}$, and that the one-row state becomes unstable
and drops from the picture, while the two-row state appears at $T=T_0^{(a)}$. The phase diagram resembles that of the ordinary
$SU(2)$ ferromagnet, with a single Curie temperature $T_0^{(a)}$ separating an unmagnetized (singlet) phase for $T>T_0^{(a)}$
and a magnetized (two-row) phase for $T<T_0^{(a)}$.

\no
However, the phase transition at $T_0^{(a)}$ is qualitatively different
from the standard one. For $T\lesssim T_0^{(a)}$ we obtain from \eqn{fhu11} the expansion for the two-row state
\be
p_1=2: \quad y =  \bigg({45 \ov 4}\bigg)^{\bb 1/4} \bigg({T_0^{(a)}-T\ov T_0^{(a)}}\bigg)^{\bb 1/4} + \dots 
\ee
from which the free energy of the two-row state obtains as
\be
 F_\text{two-row} = -{3\ov 2}  T_0^{(a)} -T \ln 6 -{2\sqrt{5} \ov 3}   {(T_0^{(a)}-T)^{3/2} \ov \sqrt{T_0^{(a)}}}
 -{25\ov 28}  {(T_0^{(a)}-T)^2 \ov T_0^{(a)}}
 +\dots \ .
\ee
The first two terms are the free energy of the singlet. Therefore, the first derivative of the free energy is continuous across
$T_0^{(a)}$ and
the transition can be characterized as a second order one, albeit the second derivative becomes infinite at
$T=T_0^{(a)}$ due to the power $3/2$ term, which is absent in the $SU(2)$ case.

\no
The truly generic cases for antisymmetric irrep start at $N=5$.

\subsubsection{Small $T$ and large $N$ limits}
For small temperature, the solution of \eqn{fhu11} for $a=p_1 /(N-p_1)$ is
\be
\label{limt0}
T\ll T_0^{(a)} : \qquad  y =  \begin{cases} {N-p_1\ov p_1} \bigg( 1 - {N\ov p_1} e^{-{N(N-1)\ov p_1 (N-2)}\, {T_0^{(s)}/ T}} +\dots  \bigg) \, , \quad p_1 > 1\ ,
\\
(N-1) \bigg( 1- N e^{-{N T_0^{(a)} \ov 2T}} +\dots  \bigg) \, , \quad  p_1 =1\  .
\end{cases}
\ee
So, for $p_1 >1$ we obtain $x_1\simeq {2\ov p_1}$ and $x_2\simeq 0$, that is, a YT with $p_1$ equal rows 
of length $ {2\ov p_1} n$. The corrections are, as in the symmetric case, nonperturbative in $T$. 
For $p_1=1$ we obtain $x_1\simeq 1$ and $\displaystyle x_2 \simeq {1\ov N-1}$, that is, a YT with a single
row of length $\displaystyle {N-2\ov N-1} n$.
We may further show that at zero temperature the  free energy becomes
\be
T\simeq 0: \qquad  F_{\rm ant} \simeq -T_0^{(a)} {N(N-1)\ov 4(N-2)}\times  \begin{cases} {4 \ov p_1}\, ,\  \ p_1 >1\ .
\\
{N\ov N-1} \, , \ \ p_1=1\ ,
\end{cases}
\label{F0}
\ee
The discontinuity in the above expressions at $p_1=1$ ensures that the free energy for the $p_1=2$
configuration is below that for $p_1=1$, consistent with the fact that, at low temperatures,
$p_1=2$ is the globally stable state.

\no
The above results can also be obtained from pure group theory. At $T=0$ the entropy term vanishes and the free energy is simply
the energy of the configuration, which is $-c/n$ times the quadratic Casimir $C^{(2)}$ of the ground state representation.
For an irrep with $p_1 \geqslant 2$ rows, the Casimir is maximized by setting the lengths of all rows to be equal to their maximal common
value, which is $2n/p_1$. For $p_1 =1$, however, due to the antisymmetry of $\chi$, we can put up to $n$ boxes in the top row,
the remaining boxes having to be distributed in the rows below. The maximum Casimir is
achieved by making the top row of maximal length $n$ and the lower $N-1$ rows of equal length $n/(N\bb-\bb1)$,
for a YT with a single row of length $n - n/(N\bb-\bb1) = n(N\bb-\bb2)/(N\bb-\bb1)$.
The Casimirs of these irreps are easily calculated and reproduce the free energies \eqn{F0}.

\no
We see that at zero temperature the YT with  $p_1=2$ rows of length $n$ has the lowest free energy, consistent
with table \ref{table:4}.

\no
Similarly, we can determine the behavior of the ferromagnet for $N\gg 1$. We obtain
\be
\begin{split}
& p_1=1:\qq y_c \simeq N \bigg( 1-{2\ov \ln N}\bigg) \ ,\qq T_c\simeq {N\ov 2 \ln N} \ ,
\\
& 
p_1>1: \qq y_c \simeq {N\ov p_1} \bigg(1-{p_1-1\ov p_1\ln N}\bigg) \ ,\qq T_c\simeq {N\ov p_1 \ln N} \ .
\end{split}
\ee

\no
In the above, only $p_1=1$ and $p_2=2$ are relevant for our considerations. 
Also, using \eqn{ta012},  the large $N$ behavior of the metastability transition temperatures from the singlet to the one- and two-row irreps 
can be found. Gathering all these together
\uu\uu\uu\be
N \gg 1: \qquad T_c^{(1)} \simeq T_c^{(2)} \simeq T_0^{(a)} {N\over 2\ln N}\ ,\qq
T_{\rm m}^{(01)} \simeq T_{\rm m}^{(02)} \simeq T_0^{(a)} {N\over 4\ln N}\ .
\ee

\no
These are similar to the corresponding results  \eqn{largeNSfun} and \eqn{largeNS} for the fundamental and symmetric irreps, but all temperatures
are smaller by a factor of two, assuming equal $T_0$'s.

\section{Conclusions }
\label{conclusions}

$SU(N)$ ferromagnets with atoms in higher irreps appear to be fascinatingly complex systems. Their
thermodynamics is, in principle, fully derivable using the formalism developed in this paper, although
the complexity of the obtained equations for higher irreps calls for a numerical (but fully doable) analysis.
Their phase structure manifests qualitatively new features compared to the case of fundamental atoms,
which include the appearance of more than two phases, and corresponding temperature regimes
in which they coexist as metastable states. In essence, such ferromagnets have more than one
ferromagnetic states with distinct polarization properties, in addition to the paramagnetic (singlet) state.

\no
At temperatures where more than one phases coexist, all but one of them are metastable. Nevertheless,
their presence is physically significant. By Arrhenius' law, the thermally driven transition of a metastable state
to a fully stable one is exponentially suppressed and the transition time is exceedingly large.
For all practical purposes, metastable states are stable if left unperturbed, and only
external perturbations (impurities, shaking the system etc.) can induce their decay. Our results, therefore,
can be physically relevant in the appropriate context.

\no
It is worth noting that, although our setting and approach involving non-itinerant atoms and
using group theory results to derive the thermodynamic properties are different than those for cold atoms,
where an effective interaction between itinerant quasiparticles leads to an effective Landau free energy
in terms of a macroscopic magnetization, some of the results are qualitatively similar, including
first-order phase transitions and metastability. The relation between our approach and the one for
cold atoms, and the possible universality of phase transitions for $SU(N)$ ferromagnetism, are
interesting questions deserving further investigation. The study of general atom irreps $\chi$
is an important tool in further probing the relation between the two systems. The possible relevance of our approach to the study of ergodicity or criticality properties of Hamiltonians based on SU(N) generators
is also a topic worth exploring.

\no
Although the analysis in this paper was done for the first two higher irreps, the doubly symmetric and
doubly antisymmetric ones, some general patterns already appear, and can be generalized here as
conjectures. Specifically, if the atom irrep $\chi$ has $r$ rows, the ferromagnet is conjectured to have
up to $r+1$ distinct phases corresponding to irreps with $0,1,\dots,r$ rows ($0$ rows corresponding to the singlet).
E.g., in the case of the $r=2$ antisymmetric irrep in this paper, $N\geqslant 5$ realizes the generic case with 3 phases, and $N=4$
the special case with 2 phases.
Moreover, the basic critical temperature $T_0^{(\chi)}$ below which the singlet becomes unstable appears
to be expressible
in terms of the quadratic Casimir of $\chi$. Specifically, its defining equation can be cast in the form 
\be
\label{cggrp}
c={T_0^{(\chi)} \ov 2} {\dim G\ov C_2^{(\chi)}}\ ,
\ee
where $c$ is the coupling constant of the model in \eqn{meanfield}, $\dim G$ is the dimension of the group
(in our case $G=SU(N)$), and  $C_2^{(\chi)}$ is the eigenvalue of the corresponding quadratic Casimir
operator. This reproduces the definitions \eqn{cnt}, \eqn{csymm}, and \eqn{cta} in this paper,
with\footnote{Expressions \eqn{c2irrepp} can be easily recovered using \eqn{lengy} and the first of \eqn{caso}.
We take $k_i=N-i$, $i=1,\dots ,N$
for the singlet, while for the fundamental $k_1=N$, for the symmetric $k_1= N+1$, and for the antisymmetric
$k_1=N$ and $k_2 = N-1$, all other $k_i$ remaining the same as for the singlet.
}
\uu\uu\uu\uu\uu
\be
\label{c2irrepp}
\begin{split}
&\dim SU(N) = N^2 -1\ , \quad C_2^{\rm fund} =  { N^2-1\ov 2 N}\ ,  \\
&C_2^{\rm sym} =  { (N-1)(N+2)\ov  N}\  ,\quad C_2^{\rm ant} =  { (N+1)(N-2)\ov  N}\ .
\end{split}
\ee
and is conjectured to hold for higher irreps.
Proving or refining the above conjectures is an interesting theoretical project.

\no
Physical applications of our ferromagnet with higher irrep $\chi$ extend to any situation where the interaction
between atoms is ferromagnetic but not fully invariant under relabeling of the states of the atoms (that is,
it is not of purely exchange type). If the coupling is exactly or approximately invariant under a
smaller unitary symmetry, the ferromagnet will be describable in the general terms of the model in this work.
Such situations are, if anything, more generic and physically relevant than the maximally symmetric
case of fundamental atom irrep, and their experimental realization is of substantial physical interest.

\no
The work in this paper generalizes the results of \cite{PSferro} in one of the possible directions suggested
there, namely higher atom irreps. Deriving the thermodynamics for irreps other than the two specific
ones worked out in this paper is an obvious and important next step. This would serve, among other purposes,
to identify the various thermodynamic states of these models and detect patterns in their transitions.
The case of the adjoint irrep is, perhaps, the most interesting one, as it is related to unpolarized atoms,
and will be examined in a future publication. Further, coupling the atoms to external
magnetic fields, for which the present paper contains the formalism, would be useful for probing
the response of the states to external fields and identifying
the full phase diagram of the model in the temperature-magnetic field space.
In view of the richness of the phase structure of such models even in the simplest case of fundamental
$\chi$ and a magnetic field in a single direction \cite{PSferro}, we expect increasingly intricate
patterns to arise for higher irreps.

\no
Other generalizations suggested in \cite{PSferro}, namely modifying the form of the two-atom interaction,
or including three- and higher-atom interactions, remain open. In a sense, higher irreps constitute a
controlled modification of the interaction, reducing the full $SU(d_\chi)$ symmetry of the atom states to
$SU(N)$, with $N<d_\chi$, but other possibilities are present. 
Higher than two-body terms would arise from higher orders in the perturbation expansion of atom interactions
and would appear as higher Casimirs of the global $SU(N)$ and/or as higher powers of Casimirs in the
mean field approximation. All such generalizations can be studied using the general formalism of this work.
A nontrivial extension of this study would be in the context of
topological phases of nonabelian models. Such topological phases have been proposed in one dimension
\cite{RQ,CFLT,TLC,RPAR} and it would be interesting to explore their existence in higher dimensions.

\no
In a different direction, the study of the
model in various large-$N$ limits is physically and mathematically interesting. In fact, the possibility to
also scale the size of the irrep $\chi$ augments the space of relevant parameters to at least three, namely
$n,N,n_\chi$ ($n_\chi$ is the number of boxes in $\chi$). Equation \eqn{lengyt} suggests the scaling
\be
N \sim n^{\alpha/2} ~,\quad  n_\chi \sim n^{1-\alpha}~,\quad  n \gg 1
\ee
for any parameter $0\leqslant \alpha \leqslant 1$. The case $N \sim \sqrt n$ studied in
\cite{PSlargeN,PSferroN} for fundamental $\chi$ ($n_\chi = 1$) corresponds to $\alpha = 1$,
but, in principle, any $\alpha \in [0,1]$
would lead to a well-defined scaling limit. The exploration of the general $\alpha$ scaling, and the
dependence of the results (if any) on the "shape" of the large $\chi$ irrep, remain intriguing topics for further
investigation.

\no
Finally, extending the model to other groups and to {\it reducible} atom irreps are other avenues
for further research. The interest of such models is that they can describe situations with more arbitrary,
or idiosyncratic, symmetry of the interactions. Other classical groups, supersymmetric groups with fermionic
dimensions, or "quantum" $q$-groups offer additional exciting possibilities.

\subsection*{Acknowledgements }

The research of A.P. was supported by the National Science Foundation 
under grant NSF-PHY-2112729 and  by PSC-CUNY grants 67100-00 55 and 6D136-00 03.\\
K.S. would like to thank the Department of Theoretical Physics at CERN, the Simons Center for Geometry and Physics at Stony Brook University and 
the Physics Department at City College of the CUNY for financial support and hospitality during the
late stages of this research. 

\appendix 

\section{Conditions for positivity}\label{appendix}

We will establish the conditions for an $N\times N$ matrix of the general form
\be
M_{ij} = f_i \delta_{ij} +\sum_{k=1}^m \e_k\, a_{k,i} a_{k,j}\ ,
\ee
to be positive definite. In the above, $f_i$ are real diagonal elements, ${\bf a}_k$, $k=1,\dots,m$ are $m$ real
$N$-component vectors, and $\e_k = \pm 1$ are the signs of the vector terms.

\no
We shall regard
$g_{ij} = f_i\, \delta_{ij}$ as a metric (of indefinite signature) under which the scalar product
of two vectors ${\bf A}, {\bf B}$ is defined as
\be
\label{inner}
{\bf A} \cdot {\bf B} = \sum_{i,j} g^{ij} A_i B_j = \sum_i f_i^{-1} A_i B_i\ .
\ee
To proceed, we assume that $s$ of the components $f_i$ are negative, which 
can always be chosen to be
$f_1,\dots,f_s <0$, the remaining $f_i$ being positive. Rescaling with the matrix $D = \diag\{1/\sqrt{|f_i|}\}$,
which preserves positivity, we bring $M$ to the form
\be
\cM_{ij} = (DMD)_{ij} = \eta_{ij}^{(s)} + \sum_{k=1}^m \e_k\, u_{k,i} u_{k,j} \ , \qq u_{k,i} ={a_{k,i}\over \sqrt{|f_i|}}\ ,
\ee
where $\eta_{ij}^{(s)}= \diag\{-1,\dots, -1; 1,\dots, 1\}$ is a flat metric with $s$ timelike
(negative metric) dimensions. The positivity of $M$, or $\cM= DMD$, depends on $s$.

\no
Positivity of $M$ requires that all eigenvalues of $\cM$ be nonnegative, or, equivalently, that $\sum_{ij} \cM_{ij} n_i n_j  >0$
for every non-vanishing vector ${\bf n}$. For the purposes of our ferromagnet calculations, only the one-vector and
two-vector cases ($m=1,2$) are relevant, and we treat them separately below.

\subsection{The one-vector case}

In principle, the one-vector case is a subcase of the two-vector case upon setting one of the vectors to zero. However,
it is instructive to work out this case explicitly. We thus consider the positivity properties of a matrix of the form
\be
\label{j9iug}
M_{ij} = f_i \delta_{ij} +\e\ a_i a_j \quad{\rm or} \quad \cM_{ij} = \eta_{ij}^{(s)} + \e\, u_{i} u_{j}\ , ~~~u_{i} ={a_{i}\over \sqrt{|f_i|}}\ .
\ee
This is relevant to the case of $\chi$ being the fundamental representation \cite{PSferro}.

\no
For $s\geqslant 2$, we can always find a vector $\bf n$ with nonzero elements
only in the timelike subspace, that is, ${\bf n} = (n_1,\dots,n_s,0,\dots,0)$, and orthogonal to $\bf u$, i.e.
${\bf u}\cdot {\bf n}  =0$, for which $\displaystyle \sum_{ij} \cM_{ij} n_i n_j  =  -\sum_{i=1}^s n_i^2 <0$ and thus the matrix is not positive definite.
Therefore, positivity is possible only for $s=0,1$:
\be
\label{stabone}
M \geqslant 0 ~~~ \Rightarrow ~~~ s=0, 1\ .
\ee
\subsubsection{The $s=0$ Euclidean case} 

For $s=0$, $\eta_{ij}^{(0)} = \delta_{ij}$ and the sign of the eigenvalues can be determined by using the $SO(N)$
invariance of the flat Euclidean metric to perform a rotation, which preserves positivity, to set the vector to $u=(u_1,0,\dots, 0)$.
Then the matrix $\cM$ becomes diagonal and we see that it has $N-1$ unit eigenvalues and a single eigenvalue
$1+\e u_1^2$, which must be positive. 
We may cast this in terms of the original vector $\bf a$ by noting that $u_1^2 = {\bf u}^2 = {\bf a}^2$,
to obtain the positivity condition
\be
\label{s0s}
\begin{split}
M\geqslant 0\ , \ \ s=0: \quad& \e= 1, \quad  {\rm any}\ {\bf a}\ ,
\\
&
\e=- 1, \quad   {\bf a}^2 <1  \ .
\end{split}
\ee

\subsubsection{The $s=1$ Minkowski case} 

We distinguish subcases depending on the norm of the vector $\bf u$. 
If $\bu$ is spacelike, then a Lorentz transformation, which preserves positivity, can set $\bu=(0,u_1,0,\dots, 0)$. 
Then clearly the component $\cM_{11}$ gives rise to a negative eigenvalue. 
If, instead, the vector $\bu$ is timelike, a Lorentz transformation can set it to $\bu=(u_1,0,\dots, 0)$. 
Then $\cM$ becomes diagonal and has $N-1$ unit eigenvalues and a single eigenvalue
$\cM_{11}=-1+\e u_1^2 $. Positivity requires $\e =1$ and $u_1^2 >1$.
Since $u_1^2 = -{\bf u}^2 = -{\bf a}^2$, we conclude that the condition for positivity is 
\be
\label{s1s}
M \geqslant 0\ , \ \ s=1 :\quad\e=1\ ,\quad {\bf a}^2 <-1 \ .
\ee

\subsection{The two-vector case}

We consider the positivity properties of a matrix involving two $N$-dimensional vectors  ${\bf a}$ and ${\bf b}$ of the form
\be\begin{split}
\label{mij}
&M_{ij} = f_i \delta_{ij} +\eo a_i a_j + \et b_i b_j\ , \quad {\rm or,} \\
&\cM_{ij} = \eta_{ij}^{(s)} + \eo u_i u_j + \et v_i v_j \ ,\quad u_i ={a_i\over \sqrt{|f_i|}}\ ,\quad v_i ={b_i\over\sqrt{|f_i|}}\ .
\end{split}\ee
As before, the positivity of $M$, or $\cM$, depends on $s$.

\no
For $s\geqslant 3$, we can always find a vector $\bf n$ with nonzero elements
only in the timelike subspace, that is, ${\bf n} = (n_1,\dots,n_s,0,\dots,0)$, and orthogonal to $\bf u$ and $\bf v$,
${\bf u}\cdot {\bf n} = {\bf v} \cdot {\bf n} =0$,
for which $\displaystyle \sum_{ij} \cM_{ij} n_i n_j  =  -\sum_{i=1}^s n_i^2 <0$ and thus the matrix is not positive definite. Thus, positivity is possible
only for $s=0,1,2$, i.e.
\be
\label{stabtwo}
M \geqslant 0 ~~~ \Rightarrow ~~~ s=0, 1, 2\ .
\ee

\subsubsection{The $s=0$ Euclidean case} 

For $s=0$, $\eta_{ij}^{(0)} = \delta_{ij}$ and, as in the one-vector case, we can perform an $SO(N)$
to set the vectors to $u=(u_1,0,\dots, 0)$ and $v=(v_1,v_2,0,\dots, 0)$.
Then the matrix $\cM$ simplifies and we see that it has $N-2$ unit eigenvalues. The remaining $2\times 2$ part of the matrix
has the form 
\be
\label{hfh111}
\cM_{2\times 2 }=\begin{pmatrix}
1+\eo u_1^2+\et v_1^2 & \et v_1 v_2  \\
\et v_1 v_2 & 1+\et v_2^2 
\end{pmatrix}\ .
\ee
We can determine the positivity of the eigenvalues of this matrix by evaluating its trace and determinant and demanding that
both be positive. These can be cast in the combinations  $u_1^2$, $v_1^2+v_2^2$ and $u_1v_1$. Expressing them in terms
of the rotationally invariant inner products ${\bf u}^2$, ${\bf v}^2$ and ${\bf u\cdot v}$, respectively, and reverting to the original vectors and metric in $M$, the conditions for positivity are
\be
\label{hf21}
 \eo {\bf a}^2 + \et {\bf b}^2 > -2 \ ,\qq  (1+\eo {\bf a}^2 )(1+ \et {\bf b}^2)> \eo \et ({\bf a}\cdot{\bf b})^2\ ,
\ee
Clearly these are automatically satisfied for $\eo=\et=1$. 
For $\e_1\e_2=-1$, the second condition implies the first one. 
For $\e_1=\e_2=-1$, the conditions are equivalent to the second one plus the requirement that one of the vectors has norm
less than 1. Summarizing,
\be
\label{s0s11}
\begin{split}
M\geqslant 0 , ~~~ s=0:\quad& \e_1=\e_2= 1:\quad  {\rm any}\ \ {\bf a} \ \ {\rm and}\ \ {\bf b}\ ,
\\
&
\e_1=-\e_2=1:\quad  ({\bf a}\cdot{\bf b})^2 > ({\bf a}^2 +1)( {\bf b}^2 -1) \ ,
\\
&
\e_1=\e_2=-1: \quad   {\bf a}^2 <1\ \ {\rm and}\ \ ({\bf a}^2 -1)({\bf b}^2 -1)>  ({\bf a}\cdot{\bf b})^2 \ .
\end{split}
\ee
Obviously for $\e_1=-\e_2=-1$ the roles of ${\bf a}$ and  ${\bf b}$ are interchanged, and in the case $ \e_1=\e_2=-1$ we
may equivalently  impose the inequality ${\bf b}^2 <1$ instead of ${\bf a}^2<1$. Note that setting ${\bf b}=0$
recovers the one-vector conditions \eqn{s0s}.

\subsubsection{The $s=1$ Minkowski  case}  

We distinguish subcases depending on the norm of the vectors.

{\bf Timelike  $\bu$:} a Lorentz transformation can bring it to the form $\bu=(u_1,0,\dots, 0)$, and a residual
$SO(N-1)$ rotation in the spacelike dimensions can bring $\bf v$ to the form 
$\bv=(v_1,v_2,0,\dots, 0)$. 
Then the matrix $\cM$ simplifies and we see that it has $N-2$ unit eigenvalues. 
The remaining part of the matrix has the form
\be
\cM_{2\times 2 }=\begin{pmatrix}
-1+\eo u_1^2+\et v_1^2 & \et v_1 v_2  \\
\et v_1 v_2 & 1+\et v_2^2 
\end{pmatrix}\ 
\ee
and, as in the $s=0$ case, its positivity can be determined by requiring its trace and determinant to be positive.
These two conditions are
\be
\label{njn11}
\e_1  u_1^ 2+ \e_2 (v_1^2+v_2^2)>0\ ,\qq -1+\e_1 u_1^2 - \e_2 (v_2^2-v_1^2)+\e_1\e_2 u_1^2 v_2^2>0\ .
\ee
We point out that, unlike $SO(N)$ rotations, Lorentz transformations are not similarity transformations and do not
preserve the eigenvalues of $\cM$. In fact, the trace is not Lorentz invariant, but the determinant is, and the determinant
condition can be expressed in the Lorentz invariant form
\be
1 +\e_1 {\bf u}^2 +\e_2 {\bf v}^2 -\e_1 \e_2 \Big(({\bf u}\cdot{\bf v})^2 -{\bf u}^2 {\bf v}^2\Big)<0 \ .
\label{trinv}\ee
 Since
Lorentz transformations preserve positivity and the trace condition is not Lorentz invariant, it must be either identically satisfied
or identically violated if the determinant condition is satisfied.
Clearly, for $\e_1=\e_2=1$ the trace condition is identically satisfied and for $\e_1=\e_2=-1$ it is not. Hence, we focus on the
cases $\e_1 \e_2=-1$.

\no
$\bullet $ $\e_1=1$ and $\e_2= -1$: if $\bv$ is timelike, it can be parametrized as $\bv= v (\cosh \phi, \sinh\phi)$.
Then the determinant condition becomes
\be
v^2 < {u_1^2-1\ov 1+ u^2_1\sinh^2\phi}\ ,\qq  u_1^2>1\ ,
\label{dtt}\ee
and the trace becomes
\be
u_1^2- v_1^2 -v_2^2 = u_1^2 - v^2\cosh 2 \phi >  {1+ (1+(1-u_1^2)^2)\sinh^2\phi \ov 1+ u^2_1\sinh^2\phi}> 0 \ ,
\ee
where we used \eqn{dtt}, and the trace condition is identically satisfied.
If $\bv$ is spacelike, it can be parametrized as
$\bv= v (\sinh \phi, \cosh\phi)$. The determinant condition becomes
\be
\begin{split}
&v^2   <  {u_1^2-1\ov u_1^2\cosh^2\phi -1} \ \ \& \ \ u_1^2>1\ , ~~~~~{\rm or}\\
&v^2  >  {u_1^2-1\ov u_1^2\cosh^2\phi -1} \ \ \& \ \ u_1^2\cosh^2\phi <  1\ .
\end{split}
\label{ddtt}\ee
Similarly, the trace becomes
\be
u_1^2- v_1^2 -v_2^2 = u_1^2 - v^2\cosh 2 \phi  \gtrless    {\sin^2\phi + (1-u_1^2)^2\cosh^2\phi \ov u^2_1\cosh^2\phi -1 }  \gtrless  0 \ ,
\ee
where we used the upper and lower cases of \eqn{ddtt} in the upper and lower inequalities. Therefore, the trace condition is
identically satisfied in the case $u_1^2 >1$ and identically violated otherwise. 

\no
$\bullet $ $\e_1=-1$ and $\e_2= 1$: if $\bv$ is spacelike, the determinant condition in \eqn{njn11} clearly cannot be satisfied.
The case $\bv$ timelike is covered by the case $\e_1=1,\e_2=-1$ upon exchanging $\bu$ and $\bv$. 
 
\no
Overall, the positivity conditions are \eqn{trinv} expressed in terms of ${\bf a},{\bf b}$ plus the additional conditions of the
above analysis; that is,
\be
\label{jfhjd}
\begin{split}
&(1+\eo \bA^2 )(1+\et \bB^2 ) < \eo \et (\ab)^2 \ ,\\
& \et = 1: \qq   {\bf a}^2 <0\ ,
\\
& \eo = -\et =1: \quad  {\bf a}^2 <-1\ .
\end{split}
\ee

\no
{\bf Spacelike $\bu$:} The case  timelike $\bv$ has been covered by the $\bu$ timelike analysis,
upon exchanging $(\e_1,\bu)$ and $(\e_2,\bv)$. For $\bv$ spacelike,
we may Lorentz transform $\bu$ and $\bv$ to $\bu = (0,u_2,0,\dots,0)$ and  $\bv = (v_1,v_2,v_3,0,\dots,0 )$.
If $|v_1| <|v_3|$ we may further Lorentz transform $\bv$ to $\bv = (0,v_2,v_3,0,\dots,0 )$.
This gives $\cM_{11}=-1$, $\cM_{1j} = \cM_{j1}=0$ ($j>1$), spoiling positivity. 
If $|v_1| >|v_3|$ we may Lorentz transform $\bv$ to $\bv = (v_1,v_2,0,\dots,0 )$. Then
\be
\cM_{2\times 2 }=\begin{pmatrix}
-1+ \et v_1^2 & \et v_1 v_2  \\
\et v_1 v_2 & 1 +  \eo u_2^2+  \et v_2^2 
\end{pmatrix}\ .
\ee
Parametrizing $\bv$ as ${\bf v}= v (\sinh\phi ,\cosh\phi)$,
and the trace and determinant conditions become
\be
\label{njn1k}
\e_1  u_2^ 2+ \e_2 v^2 \cosh2\phi>0\ ,\qq 1+\e_1 u_2^2 + \e_2 v^2-\e_1\e_2 v^2 u_2^2 \sinh^2\phi<0\ .
\ee
For $\e_1\bb=\e_2\bb=\bb1$ the trace condition is identically satisfied and for $\e_1\bb=\e_2\bb=\bb-1$ it is not. 
Hence, we will examine only the cases $\e_1\e_2 = -1$.

\no
$\bullet$ $\e_1=1$ and $\e_2= -1$: the determinant condition becomes
\be
v^2 >{1+u_2^2\ov 1-u_2^2\sinh^2\phi}\ ,\qq u_2^2\sinh^2\phi<1\ ,
\label{sss}\ee
and the trace becomes
\be
u_2^2 - v^2\cosh 2 \phi < - {\cosh^2\phi+ (1+u_2^2)^2\sinh^2\phi \ov 1-u^2_2\sinh^2\phi  }< 0 \ ,
\ee
upon using \eqn{sss}, and therefore the trace condition cannot be satisfied.

\no
$\bullet $  $\e_1=-1$ and $\e_2= 1$: this is covered by exchanging $\bu$ and $\bv$ above.

\no
Altogether, the positivity conditions expressed in the original vectors are
\be
\label{jfhjh44c}
\begin{split}
M \geqslant 0, \quad s=1:\quad & (1+\eo \bA^2 )(1+\et \bB^2 ) < \eo \et (\ab)^2\ ,
\\
& (\eo,\et ) =(1,1):\qq   \ {\rm any}\  {\bf a}\, , {\bf b}\ ,
\\
&
(\eo,\et ) =(1,-1): \quad  {\bf a}^2 < -1\ . 
\\
\end{split}
\ee
The case $(\eo,\et ) =(-1,1)$ is covered by interchanging $\bf a$ and $\bf b$.
Note that setting ${\bf b}=0$ recovers the conditions \eqn{s1s} for the one-vector case.

\subsubsection{The $s=2$ case}

If either of $\bu$ or $\bv$ is spacelike, say $\bu$, a generalized Lorentz transformation can bring it to the
form $\bu = (0,0,u_3,0,\dots,0)$ while a rotation within the space dimensions and a rotation within the two time dimensions
can achieve $\bv = (0,v_2,v_3,v_4,0,\dots,0)$. 
Then $\cM_{11} =-1$, and the matrix is not positive. Therefore, both $\bu$ and $\bv$ must be timelike.
In this case, we can set $\bu = (u_1,0,\dots,0)$ and
 $\bv = (v_1,v_2,v_3,0,\dots ,0)$. With further rotations, we may set 
 $\bv = (v_1,0,v_3,0,\dots ,0)$ if $|v_2|<|v_3|$ or $\bv = (v_1,v_2,0,\dots ,0)$ if $|v_2|>|v_3|$.
The former case leads to $\cM_{22} =-1$ and the matrix is not positive. The latter case gives
\be
\cM_{2\times 2 }=\begin{pmatrix}
-1+ \eo u_1^2 +  \et v_1^2   & \et v_1 v_2  \\
\et v_1 v_2 & -1 +   \et v_2^2 
\end{pmatrix}\ .
\ee
Parametrizing $\bv$ as as $\bv= v (\cos \phi, \sin\phi)$, the trace and determinant conditions are
\be
\label{njn13}
\e_1  u_1^ 2+ \e_2 v^2 >2\ ,\qq 1-\e_1 u_1^2 - \e_2 v^2 +\e_1\e_2 u_1^2 v^2 \sin^2\phi >0\ .
\ee
As in the $s=1$ case, the trace is not Lorentz invariant, while the determinant is.
Clearly for $\e_1=\e_2=-1$ the trace condition cannot be satisfied. Hence, we examine the other three cases.

\no
$\bullet $ $\e_1=1$ and $\e_2= 1$: the determinant condition gives
\be
v^2  <  {1-u_1^2\ov 1- u^2_1\sin^2\phi}\ ,\quad  u_1^2<1 \qq {\rm or}\qq  v^2  >  {1-u_1^2\ov 1- u^2_1\sin^2\phi}\ ,
\quad  u^2_1\sin^2\phi>1\ . 
\ee
In the first case the trace becomes
\be
u_1^2 + v^2 -2 < u_1^2-2 +{1-u_1^2\ov 1-u_1^2 \sin^2\phi} < u_1^2-1<0 \ ,
\ee
and the trace condition is not satisfied. In the second case the trace becomes
\be
u_1^2 + v^2 -2 > u_1^2-2 +{1-u_1^2\ov 1-u_1^2 \sin^2\phi} > u_1^2-1 > 0 \ ,
\ee
and the trace condition is satisfied.

\no
$\bullet$ $\e_1=-1$ and $\e_2= 1$: the determinant condition gives
\be
v^2  <   {1+u_1^2 \ov 1+ u^2_1\sin^2\phi}\ .
\ee
The trace becomes
\be
-u_1^2 + v^2 -2 < -u_1^2-2 +{1+u_1^2\ov 1+u_1^2 \sin^2\phi} < -1 
\ee
and the trace condition is not satisfied.

\no
$\bullet $ $\e_1=1$ and $\e_2=- 1$: recovered by interchanging $\bu$ and $\bv$ in the previous case. 

\no
Expressing $u_1^2$, $v^2$ and $u_1v_1$ in terms of the Lorentz invariant inner
products $-{\bf u}^2$, $-{\bf v}^2$ and $-{\bf u\cdot v}$, respectively, the positivity conditions in terms of the
original vectors are
\be
\label{fjh31}
M \geqslant 0\ ,\quad s=2:\quad \eo  = \et=1\ ,\quad \bA^2<-1\ ,\quad
(1+ {\bf a}^2 )(1+  {\bf b}^2 ) > ({\bf a}\cdot {\bf b})^2 \ .
\ee
The condition
$({\bf a}\cdot {\bf b} )^2 < {\bf a}^2 {\bf b}^2$ required for $|v_2| > |v_3|$ is a trivial consequence of \eqn{fjh31}.

\end{document}